\documentclass[11pt,a4paper]{article}
\pdfoutput=1

\usepackage{jheppub}

\usepackage{mathrsfs}
\usepackage{amsmath,amsthm,amssymb,slashed}
\usepackage{color}
\usepackage{multirow}
\usepackage[normalem]{ulem}
\usepackage{afterpage}

\newcommand{\be}{\begin{equation}}
\newcommand{\ee}{\end{equation}}
\newcommand{\bea}{\begin{eqnarray}}
\newcommand{\eea}{\end{eqnarray}}
\newcommand{\ba}{\begin{array}}
\newcommand{\ea}{\end{array}}

\allowdisplaybreaks

\title{New physics from oscillations at the DUNE near detector, and the role of systematic uncertainties
}

\author[a]{Pilar Coloma,}
\affiliation[a]{Instituto de Física Teórica UAM/CSIC, Calle de
  Nicolás Cabrera 13--15, Universidad Autónoma de Madrid,
  Cantoblanco, E-28049 Madrid, Spain}
\emailAdd{pilar.coloma@ift.csic.es}

\author[b]{Jacobo L\'opez-Pav\'on,}
\affiliation[b]{Instituto de Física Corpuscular, Universidad de Valencia \& CSIC, 
Edificio Institutos de Investigaci\'on, Calle Catedr\'atico Jos\'e Beltr\'an 2, 46980 Paterna, Spain}
\emailAdd{jacobo.lopez@uv.es}

\author[a,c]{Salvador Rosauro-Alcaraz,}
\emailAdd{salvador.rosauro@uam.es}
\affiliation[c]{Departamento de F\'isica Te\'orica, 
Universidad Aut\'onoma de Madrid,Campus de Cantoblanco, 28049 Madrid, Spain}

\author[b]{Salvador Urrea}
\emailAdd{salvador.urrea@ific.uv.es}

\abstract{
We study the capabilities of the DUNE near detector to probe deviations from unitarity of the leptonic mixing matrix, the 3+1 sterile formalism and Non-Standard Interactions affecting neutrino production and detection. We clarify the relation and possible mappings among the three formalisms at short-baseline experiments, and we add to current analyses in the literature the study of the $\nu_\mu \to \nu_\tau$ appearance channel. We study in detail the impact of spectral uncertainties on the sensitivity to new physics using the DUNE near detector, which has been widely overlooked in the literature. Our analysis shows that this plays an important role on the results and, in particular, that it can lead to a strong reduction in the sensitivity to sterile neutrinos from $\nu_\mu \to \nu_e$ transitions, by more than two orders of magnitude. This stresses the importance of a joint experimental and theoretical effort to improve our understanding of neutrino nucleus cross sections, as well as hadron production uncertainties and beam focusing effects. Nevertheless, even with our conservative and more realistic implementation of systematic uncertainties, we find that an improvement over current bounds in the new physics frameworks considered is generally expected if spectral uncertainties are below the $5\%$ level.
}

\preprint{FTUV-21-0505.4636, IFT-UAM/CSIC-21-51, IFIC/21-14}

\keywords{sterile neutrinos, neutrino oscillations, unitarity, neutrino mixing }

\begin{document}

\maketitle

\section{Introduction}

The neutrino oscillation picture is still far from being complete. While T2K~\cite{T2K:nu2020} and NOvA~\cite{NOvA:nu2020} will keep collecting data in the coming years, new facilities will be needed to test at high confidence level whether CP is violated in the leptonic sector, and to determine the neutrino mass ordering with a statistical significance above the 5$\sigma$ level. For that purpose, before the end of the decade two major long-baseline neutrino oscillation experiments will go online: the Deep Underground Neutrino Experiment~\cite{Abi:2020qib} (DUNE) in the US, and the Tokai-to-Hyper-Kamiokande~\cite{Abe:2018uyc,Abe:2015zbg} (T2HK) experiment in Japan. 

With their larger detectors and more powerful beams, the upcoming generation of oscillation experiments will also be able to thoroughly test the robustness of the standard three-family oscillation framework at an unprecedented level of precision. In fact, while there is plenty of evidence indicating that at least two of the Standard Model (SM) neutrinos are massive, the exact mechanism responsible for their generation is still unknown and requires the addition of new degrees of freedom to the SM particle content, which may lead to the observation of new physics effects. The most minimal extension that can generate neutrino masses is the inclusion of Majorana right-handed neutrinos through the celebrated type-I Seesaw mechanism~\cite{Minkowski:1977sc,Mohapatra:1979ia,Yanagida:1979as,GellMann:1980vs} which leads to unitarity deviations for the Pontecorvo-Maki-Nakagawa-Sakata (PMNS) $3\times3$ active-light sub-block of the full neutrino mixing matrix and, depending on the mass scale, to additional oscillation frequencies in the probabilities~\cite{deGouvea:2005er,Liao:2006rn,deGouvea:2006gz,Donini:2011jh,Donini:2012tt}. Alternatively, following a completely model-independent perspective, one could add higher-dimensional effective operators to the SM Lagrangian stemming from a new physics theory at high energies. Such operators may induce new interactions affecting neutrino production, propagation or detection processes, usually referred to as Non-Standard neutrino Interactions (NSI)~\cite{Wolfenstein:1977ue,Mikheev:1986gs,Roulet:1991sm,Guzzo:1991hi,Grossman:1995wx,Bergmann:1999rz}. 

It is clear that, in order to search for new physics, systematic uncertainties should be reduced as much as possible. At long-baseline neutrino oscillation experiments, these are relatively large and stem from two key limitations: (i) our inability to compute the neutrino flux analytically; and (ii) the lack of a microscopic model describing the neutrino-nucleus interaction cross section at GeV energies, where nuclear effects are very relevant. In order to reduce these, long-baseline experiments typically use a near detector (ND), located at a sufficiently short baseline (typically $L_{\rm ND}\sim \mathcal{O}(500)$ m) so that standard neutrino oscillations have not yet developed. The ND  data is used to determine the unoscillated spectrum with high accuracy, and its combination with the far detector (FD) data leads to a partial cancellation of systematic uncertainties (see, e.g., Ref.~\cite{Alvarez-Ruso:2017oui} for a review on the theoretical and experimental challenges that have to be faced for this to take place).

A priori, the ND data alone can be used to constrain several of the new physics scenarios outlined above and, in some cases, can provide complementary information to that available at the FD. This is clear in the case of an eV-scale sterile neutrino: for example, at DUNE the new oscillation frequency would roughly match the ND distance and induce an oscillating pattern~\cite{Choubey:2016fpi,Tang:2017khg,Miranda:2018yym}, while at the FD this effect would be completely averaged-out. In the case of NU of the mixing matrix (or, equivalently, NSI in production and detection), the ND data can be used to probe parameters which would otherwise essentially be inaccessible at the FD due to its much lower statistics. Remarkably, this type of new physics effects can induce a neutrino flavor transition already at zero distance. Although a zero-distance effect in the context of non-unitarity of the PMNS matrix has been studied in the context of Neutrino Factories~\cite{Antusch:2006vwa,Goswami:2008mi,Antusch:2009pm,Meloni:2009cg,Antusch:2010fe}, to the best of our knowledge this issue has not been explored for the DUNE ND, with the notable exception of Ref.~\cite{Miranda:2018yym} where different configurations for the ND complex were considered. Our work differs from the study in Ref.~\cite{Miranda:2018yym} in several aspects. In particular, we make use of the latest DUNE configuration details in our simulations (beam configuration, detector efficiencies and resolutions, etc). Additionally, while in Ref.~\cite{Miranda:2018yym} the zero-distance effect was only considered for the $\nu_\mu \to\nu_e$ channel, our analysis also includes a study of the sensitivity to anomalous $\nu_\tau$ appearance in the ND.  While some $\nu_\tau$ would in principle be produced at the target from the decay of heavy mesons and $\tau$ leptons, the flux at DUNE would be too small to lead to an observable number of events in the ND. Thus, the observation of a $\nu_\mu \to \nu_\tau$ signal at the ND would automatically point to new physics.

However, while the ND can be used on its own, from the reasoning above it also becomes clear that it will be affected by a much larger level of systematic uncertainties than FD data. In fact, while some of these uncertainties will be correlated among different energies (leading to an overall ``rescaling'' effect for the event rates), in some cases they can lead to modifications of the expected shape of the event spectrum. Therefore, when studying the sensitivity of neutrino NDs to new physics it is important to distinguish between normalization and shape uncertainties, as these will affect the sensitivity to experimental observables differently. It is however remarkable that most studies of the sensitivity to new physics using the ND at long-baseline experiments only include normalization uncertainties. In this work, we study in detail the impact of shape uncertainties on the sensitivity of the DUNE ND to three main new physics scenarios: non-unitarity (NU) of the PMNS mixing matrix, sterile neutrino oscillations, and Non-Standard Interactions (NSI) affecting neutrino production and detection processes. In particular, we will quantify the expected loss in sensitivity when shape uncertainties are included in the analysis, which stresses the importance of keeping these under control. In this sense, our approach is similar to the one adopted in Refs.~\cite{Miranda:2018yym, Meloni:2018xnk}. However, note that in Ref.~\cite{Meloni:2018xnk} the focus was put on the far detector analysis at DUNE and the sensitivity to NSI in neutrino propagation, while in Ref.~\cite{Miranda:2018yym} the authors considered a different set of near detector configurations (baseline, energy resolution, etc). 

The paper is organized as follows. Section~\ref{sec:notation} presents the formalism used and clarifies the relation and possible mapping between the NSI, NU and sterile neutrino formalisms for short-baseline experiments. The most important simulation details, event rates and our choice of systematic uncertainties is discussed in Sec.~\ref{sec:sim}, while our results are presented in Sec.~\ref{sec:results}. We summarize and conclude in Sec.~\ref{sec:conclusions}. Appendices~\ref{app:events},~\ref{app:chi2} and~\ref{app:average-out} include more technical details.

\section{Theoretical framework and notation}
\label{sec:notation}

As mentioned above, the simplest extension able to account for the light neutrino masses and leptonic mixing observed in neutrino oscillation experiments, consists in the addition of singlet fermions to the SM field content. The light neutrino masses and mixing can be successfully generated via the seesaw mechanism, or its different low scale realizations as the inverse and direct seesaw models~\cite{Mohapatra:1986aw,Mohapatra:1986bd,Bernabeu:1987gr,Branco:1988ex,Malinsky:2005bi}, for masses of the new fermions in the range between the $\mathcal{O}({\rm eV})$ and up to (near) the GUT scale. These new states are typically named in the literature as right-handed neutrinos or heavy neutral leptons (HNLs) when their mass lies around or above the electroweak scale, and sterile neutrinos when this new physics scale is close to the $\mathcal{O}({\rm eV})$ region as relevant for the LSND~\cite{Aguilar:2001ty}, MiniBooNE~\cite{Aguilar-Arevalo:2020nvw} and reactor anomalies~\cite{Mention:2011rk,Huber:2011wv} (see e.g., Refs.~\cite{Dentler:2017tkw,Dentler:2018sju,Moulai:2019gpi,Giunti:2020uhv,Berryman:2019hme,Gariazzo:2018mwd} for recent global fits in the context of light sterile neutrinos).

In all generality, the full mixing matrix connecting the mass and flavor basis (including light and heavy states) can be written as
\be
\label{eq:U}
\mathcal{U} = \begin{pmatrix}
N & \Theta  \\
R & S \\
\end{pmatrix} \, ,
\ee
Here $N$ is a $3\times3$ non-unitary matrix corresponding to the PMNS active-light sub-block, and $\Theta$ is the $3\times n$ sub-block parameterizing the mixing between active and heavy neutrinos, with $n$ the number of new states. The $R$ ($S$) sub-block gives the mixing between the sterile and light (heavy) states and does not play any role in neutrino oscillations. 

The phenomenology strongly depends on the mass scale of the new states. According to their impact on neutrino oscillations, we can essentially distinguish two possible scenarios that we will describe below in more detail: $(i)$ NU generated by new states above the electroweak scale, which can be integrated out from the low energy spectrum so that the low energy new physics effects are directly encoded in the active-light PMNS sub-block $N$ (which is no longer unitary); and $(ii)$ kinematically accessible sterile neutrinos, which can be produced in the neutrino beam and lead to additional oscillation frequencies in the probabilities. 

Notice that even in the sterile neutrino case, the $N$ sub-block is not unitary and thus this scenario can be considered as a source of deviations from unitarity from low scales. Indeed, in \cite{Blennow:2016jkn} it was shown that when the new squared-mass differences are large enough as compared with the ratio between the experiment's baseline and energy, the effects in long-baseline neutrino oscillations coincide \emph{at leading order} (linear on the parameters quantifying the deviations from unitarity) with those generated in the NU scenario stemming from heavy scales. In such a case, an averaged-out limit in which the oscillation probabilities become independent of the new frequencies is achieved and the effects in the far detector are virtually equivalent for both cases. As we will show below, for the appearance channels in short-baseline experiments it is necessary to go beyond the linear order, which results into a factor of two difference in the oscillation probabilities between the two scenarios in this case.

\subsection{Parameterization}

The most popular parameterizations of the non-unitary matrix $N$ are given by 
\be
N  = (I-T) U   \quad \mathrm{or} \quad N =(I-\eta) U',
\label{eq:N}
\ee
where $\eta$ is an hermitian matrix~\cite{Broncano:2002rw,FernandezMartinez:2007ms}, while $(I-T)$ is a lower triangular matrix~\cite{Xing:2007zj,Xing:2011ur,Escrihuela:2015wra,Li:2015oal} given by\footnote{Notice that in~\cite{Escrihuela:2015wra} the subindices for the $\alpha$ parameters are indicated by numbers while we use flavor indices. In addition, here $\alpha_{\beta\beta}$ are small parameters which directly parameterize the deviations from unitarity while in~\cite{Escrihuela:2015wra} $\alpha_{jj}$ are close to one, in such a way that $\alpha_{jj}$=1-$\alpha_{\beta\beta}$.}
\be
T = \begin{pmatrix}
\alpha_{ee} & 0 & 0\\
\alpha_{\mu e} & \alpha_{\mu \mu} & 0\\
\alpha_{\tau e} & \alpha_{\tau \mu} & \alpha_{\tau \tau}
\end{pmatrix},
\label{eq:alpha}
\ee
and $U$ and $U'$ are unitary matrices that resemble the standard PMNS matrix up to small corrections driven by $\alpha$ and $\eta$. Both parameterizations are equally general and a mapping between them can be found in~\cite{Blennow:2016jkn}. In this work we will use the lower triangular parametrization but our results can thus be trivially mapped to the hermitian parameterization.

As we will see, our analysis can be performed in terms of the $\alpha$ parameters for both scenarios, NU sourced from heavy new physics and sterile neutrino oscillations. In the literature, the sterile neutrino analyses are typically done as a function of the mixing between the active and heavy (mostly sterile) states. The connection between both formalisms can be easily done just considering the unitarity of the full mixing matrix $\mathcal{U}\mathcal{U}^\dagger=I$, which implies the following simple relation between the $N$ matrix and the active-heavy mixing:
\be
NN^\dagger+\Theta\Theta^\dagger=I=I-T-T^\dagger+\Theta\Theta^\dagger+\mathcal{O}(\alpha^2),
\ee
where in the right hand side we have introduced the triangular parameterization for $N$ given by Eqs.~(\ref{eq:N}) and (\ref{eq:alpha}). From this equation we arrive to
\bea
\alpha_{\beta\beta}&=&\frac{1}{2}\left(\Theta\Theta^\dagger\right)_{\beta\beta}=\frac{1}{2}\sum_{i=4}^n |\mathcal{U}_{\beta i}|^2,\nonumber\\
\alpha_{\gamma\beta}&=& \left(\Theta\Theta^\dagger\right)_{\gamma\beta} =\sum_{i=4}^n \mathcal{U}_{\gamma i}\mathcal{U}_{\beta i}^*.
\label{eq:alphavsU}
\eea

Even though both scenarios can be studied with the same formalism using the same parameterization, the bounds applying to each case are remarkably different. If the deviations from unitarity are generated from very heavy scales, integrating out the heavy states leads to modifications of the charged-current and neutral-current couplings of the active neutrinos and very stringent constraints are derived from precision electroweak and flavor searches~\cite{Langacker:1988ur,Nardi:1994iv,Tommasini:1995ii,Antusch:2006vwa,Alonso:2012ji,Akhmedov:2013hec,Antusch:2014woa,Fernandez-Martinez:2015hxa,Fernandez-Martinez:2016lgt}. On the other hand, when the new states are light enough to be kinematically produced with negligible masses as compared with the energy scale of these experiments, unitarity is effectively restored in those observables and the just mentioned constraints do not apply. In such a case, the present bounds stem from neutrino oscillation experiments and are less stringent. A summary of the current bounds in both scenarios is shown in Tab.~\ref{tab:bounds}, extracted from Ref.~\cite{Blennow:2016jkn}. Note that the constraints in the middle column apply for new squared-mass differences associated to the sterile neutrinos in the range $\Delta m^2 \gtrsim 100$~eV$^2$, and are thus relevant for both near and far detectors of most long-baseline experiments when the sterile neutrino oscillations are in the averaged-out regime. The bounds shown in the right column apply for $\Delta m^2 \sim 0.1-1$~eV$^2$, being relevant if the sterile neutrino oscillations are in the averaged-out regime only in the far detector.

\begin{table}[t!]
\setlength{\tabcolsep}{7pt}
\begin{center}
\renewcommand{\arraystretch}{1.6}
\begin{tabular}{|  c@{\quad} | c@{\quad} | c@{\quad} c@{\quad}   | }
\hline
    & High-scale Non-Unitarity & \multicolumn{2}{c|}{Low-scale Non-Unitarity} \\
     & ($m>$ EW)  &  $\Delta m^2 \gtrsim 100$~eV$^2$ & $\Delta m^2 \sim 0.1-1$~eV$^2$\\ \hline\hline
$\alpha_{ee} $ & $1.3 \cdot 10^{-3}$~\cite{Fernandez-Martinez:2016lgt} & $2.4 \cdot 10^{-2} $~\cite{Declais:1994su} & $1.0 \cdot 10^{-2} $~\cite{Declais:1994su}\\
$\alpha_{\mu\mu}$ & $2.2 \cdot 10^{-4}$~\cite{Fernandez-Martinez:2016lgt} & $2.2 \cdot 10^{-2}$~\cite{Abe:2014gda} & $1.4 \cdot 10^{-2}$~\cite{MINOS:2016viw}\\
$\alpha_{\tau\tau}$ & $2.8 \cdot 10^{-3}$~\cite{Fernandez-Martinez:2016lgt} & $1.0 \cdot 10^{-1}$~\cite{Abe:2014gda} & $1.0 \cdot 10^{-1}$~\cite{Abe:2014gda}\\
$\lvert\alpha_{\mu e}\rvert$ & $6.8 \cdot 10^{-4} \; (2.4 \cdot 10^{-5})$~\cite{Fernandez-Martinez:2016lgt} & $2.5 \cdot 10^{-2} $~\cite{Astier:2003gs} & $1.7 \cdot 10^{-2} $ \\
$\lvert\alpha_{\tau e}\rvert$ & $2.7 \cdot 10^{-3}$~\cite{Fernandez-Martinez:2016lgt} & $6.9 \cdot 10^{-2}$ & $4.5 \cdot 10^{-2}$ \\
$\lvert\alpha_{\tau\mu}\rvert$ & $1.2 \cdot 10^{-3}$~\cite{Fernandez-Martinez:2016lgt} & $1.2 \cdot 10^{-2}$~\cite{Astier:2001yj} & $5.3 \cdot 10^{-2}$ \\ \hline\hline
\end{tabular}
\caption{\label{tab:bounds} Upper bounds on the Non-Unitarity (NU) framework using the $\alpha$ parametrization, extracted from Ref.~\cite{Blennow:2016jkn}. The constraints are shown at $2\sigma$ and 95\% CL (1 d.o.f.) for NU stemming from very heavy scales and low scale physics (averaged-out light sterile neutrinos) respectively. The value quoted in parenthesis for the $\alpha_{\mu e}$ element corresponds to the bound obtained from $\mu \to e \gamma$. The limits for the off-diagonal parameters without a reference are derived indirectly from bounds on the diagonal parameters via $|\alpha_{\alpha \beta}| \leq 2 \sqrt{\alpha_{\alpha \alpha} \alpha_{\beta \beta}}$. See~\cite{Blennow:2016jkn} for further details.}
\end{center}
\end{table}

\subsection{Non-Unitarity from new physics above the electroweak scale}
\label{sec:notation-NU}

In this scenario the heavy states are integrated out from the low energy spectrum and are thus not kinematically accessible in the experiment. We will focus our analysis on the sensitivity provided by the DUNE ND complex and, therefore, we will restrict our study to neutrino oscillations in vacuum\footnote{See for instance~\cite{Blennow:2016jkn} for a neutrino oscillation analysis including matter effects.}. The associated oscillation probability is given by~\cite{Antusch:2006vwa,Blennow:2016jkn}
\begin{equation}
\label{eq:PNU}
P_{\gamma\beta} = \left\lvert \sum_j N_{\beta j} N^*_{\gamma j}\,e^{\frac{-i\Delta m^2_{j1}L}{2E}}\right\rvert^2 \, ,
\end{equation}
where $L$ and $E$ correspond to the baseline and the neutrino energy, respectively. Notice that this theoretical probability should be convoluted with the neutrino flux and cross section in order to obtain the number of events. 

At the experimental level, in a far detector analysis the oscillation probability is obtained from the ratio between the number of events observed in the FD and an extrapolation of the results from the ND. Taking into account the corrections from NU, this results in the following experimentally inferred oscillation probability~\cite{Blennow:2016jkn}
\begin{equation}
\label{eq:PNUexp}
\mathcal P_{\gamma\beta} = \frac{P_{\gamma\beta}}{((NN^\dagger)_{\gamma\gamma})^2}.
\end{equation}
The normalization factor can play a relevant role in long-baseline neutrino oscillation studies~\cite{Blennow:2016jkn}, which is often overlooked in the literature. 

However, at very short distances, as the ones considered in this work, the theoretical and experimentally inferred oscillation probabilities for the appearance channels coincide. In particular, we get\footnote{Notice that a normalization factor as the one present in Eq.~(\ref{eq:PNUexp}) would not play any role here since it would only give a subleading contribution to Eq.~(\ref{eq:PNUL0}).}:
\bea
\label{eq:PNUL0}
P_{\mu e}(L=0) &=& |\alpha_{\mu e}|^2,\nonumber\\
P_{\mu\tau}(L=0) &=& |\alpha_{\tau\mu}|^2.
\eea
From this equation it is clear that, if deviations from unitarity are induced by new physics above the electroweak scale, there is a non-zero probability of observing flavor transitions already at zero distance. This renders the neutrino NDs as a powerful tool to constrain or potentially discover a new physics signal encoded in $\alpha_{\gamma\beta}$. 

\subsection{Sterile neutrinos \& Non-Unitarity from new physics at low scales}
\label{sec:notation-sterile}

If the new states are light enough to be kinematically produced in the neutrino beam, they can directly participate in the oscillations. Therefore, a priori, the oscillation phenomenology is expected to be different from the NU case considered above. However, it has already been shown that in the average-out limit both scenarios can be virtually equivalent at the phenomenological level regarding the far detector physics~\cite{Blennow:2016jkn}. Here we will focus on the ND physics paying particular attention to the differences or similarities between the two approaches. For simplicity, for the sterile neutrino framework we will consider the $3+1$ case in which only one extra sterile neutrino is added.

At very short baselines (as is the case for the DUNE ND), the three-family oscillation frequencies are very suppressed and the oscillation probabilities can be written as 
\bea
P_{\gamma\beta}&=& 4\,|\mathcal{U}_{\beta 4}|^2|\mathcal{U}_{\gamma 4}|^2\sin^2\left(\frac{\Delta m^2_{41}L}{4E}\right), \nonumber\\
P_{\beta\beta}&=& 1-4\,|\mathcal{U}_{\beta 4}|^2(1-|\mathcal{U}_{\beta 4}|^2)\sin^2\left(\frac{\Delta m^2_{41}L}{4E}\right),
\label{eq:Psteriles}
\eea
to a very good approximation. It is common in the literature to parametrize these probabilities in terms of effective mixing angles for the appearance and disappearance channels, as
\bea
\label{eq:sterile-eff-angles}
P_{\gamma\beta} =& 
 \sin^22\vartheta_{\gamma\beta}\sin^2\left(\dfrac{\Delta m^2_{41}L}{4E}\right), \quad & \rm{with} \;\; \sin^22\vartheta_{\gamma\beta} \equiv 4\,|\mathcal{U}_{\beta 4}|^2|\mathcal{U}_{\gamma 4}|^2 , \nonumber\\
P_{\beta\beta} = & 1 - \sin^22\vartheta_{\beta\beta}\sin^2\left(\dfrac{\Delta m^2_{41}L}{4E}\right), \quad & \rm{with} \;\; \sin^2\vartheta_{\beta\beta} \equiv |\mathcal{U}_{\beta 4}|^2 \, .
\eea 
In the averaged-out regime ($\Delta m^2_{41} L/E \gg 1$) the oscillations are too fast to be distinguished at the detector\footnote{See appendix~\ref{app:average-out} for details on the numerical implementation of this regime in the simulations.}, and we find
\bea
\label{eq:PsterilesL0_1}
P_{\mu e}&=& 2\,|\mathcal{U}_{\mu 4}|^2|\mathcal{U}_{e 4}|^2 = 2|\alpha_{\mu e}|^2,\\
\label{eq:PsterilesL0_2}
P_{\mu\tau}&=& 2\,|\mathcal{U}_{\mu 4}|^2|\mathcal{U}_{\tau 4}|^2 = 2|\alpha_{\tau\mu}|^2,\\
\label{eq:PsterilesL0_3}
P_{\beta\beta}&=& 1-2\,|\mathcal{U}_{\beta 4}|^2(1-|\mathcal{U}_{\beta 4}|^2) = 1 - 4\alpha_{\beta\beta}+\mathcal{O}\left(\alpha^2\right).
\eea
where in the right-hand side we have just introduced Eq.~(\ref{eq:alphavsU}) particularized to the $n=1$ case under consideration. 

In the absence of an auxiliary detector with an even closer location to the neutrino source, the ND has a sensitivity to the diagonal $\alpha_{\beta\beta}$ parameters through disappearance channels which is ultimately determined by the normalization uncertainty. Due to the limited knowledge of the neutrino flux and cross section, the region of $\alpha_{\beta\beta}$ that can be experimentally probed by the ND is already ruled out by present experiments. Therefore, in this regime, only the appearance channels are relevant to constrain the new physics effects with the ND. 

Furthermore, we would like to stress that the difference between Eq.~(\ref{eq:PNUL0}) and Eqs.~(\ref{eq:PsterilesL0_1})-(\ref{eq:PsterilesL0_2}) is only a global factor 2. This means that the effects at the ND produced by NU stemming from heavy new physics and averaged-out sterile neutrinos are practically indistinguishable. It is potentially possible to distinguish if NU effects measured in the ND are generated by heavy or light new physics~\cite{Fong:2016yyh} if complementary observables are added to the analysis. However, when the corrections are sourced by new physics above the electroweak scale, the current constraints from other observables become very stringent (see Tab.~\ref{tab:bounds}) and essentially exclude the possibility of observing any signal at DUNE. Therefore, we will focus on the case in which the deviations from unitarity are generated by new physics at low scales (i.e., sterile neutrinos in the average-out limit). For brevity, in the rest of this work we will refer to this scenario simply as NU. In any case, our results can be trivially recasted to the heavy NU case rescaling our sensitivity limits by the corresponding factor of two.

\subsection{NSI}
\label{sec:notation-NSI}

The Non-Standard neutrino Interaction (NSI) framework is a model-independent phenomenological approach in which the new physics effects in neutrino oscillation experiments are parameterized in terms of four-fermion effective operators (see e.g. Refs.~\cite{Falkowski:2019xoe,Falkowski:2019kfn} for a QFT description of NSI interactions in the context of SMEFT and reactor experiments).  While obtaining sizable NSI at low energies from new physics at high energies without entering in conflict with charged lepton processes is challenging~\cite{Gavela:2008ra, Antusch:2008tz}, it has been shown in recent literature that a possible way out of this argument is the inclusion of new degrees of freedom at low scales~\cite{Farzan:2015doa,Farzan:2015hkd,Farzan:2016wym,Babu:2017olk} or in radiative models of neutrino masses~\cite{Babu:2019mfe} (see~\cite{Dev:2019anc} for a recent review on the viability of NSI models in this context). In this work we will follow a purely phenomenological approach and report the expected bounds on NSI without considering the underlying theory that can lead to the four-fermion effective operators at low energies.

NSI would lead to corrections in neutrino production, neutrino detection, or in neutrino propagation through matter. Since we are considering the ND complex, there are no effects in neutrino propagation and only NSI in production and detection will be discussed here. The appearance oscillation probability ($\gamma\neq\beta$) in presence of NSI can be parameterized as
\be
\label{eq:prob-NSI}
P_{\gamma\beta}(L=0)=\left\lvert\left[(I+\epsilon^d)\left(I+\epsilon^s\right)\right]_{\beta\gamma}\right\rvert^2
=|\epsilon^d_{\beta\gamma}|^2+|\epsilon^s_{\beta\gamma}|^2+2|\epsilon^d_{\beta\gamma}||\epsilon^s_{\beta\gamma}|\cos(\Phi^s_{\beta\gamma}-\Phi^d_{\beta\gamma}),
\ee
where $\epsilon^{s(d)}_{\beta\gamma}=|\epsilon^{s(d)}_{\beta\gamma}|e^{i\Phi^{s(d)}_{\beta\gamma}}$. It is clear that, analogously to the NU case, a flavor transition can already occur at zero distance in presence of NSI. The main difference between these two approaches at the phenomenological level is that production and detection are correlated in the NU framework, while this is not generally the case in the NSI scenario. Indeed, NU can be considered a particular case of NSI in production and detection. In particular, comparing the above equation with Eqs.~(\ref{eq:PsterilesL0_1})-(\ref{eq:PsterilesL0_2}) it is straightforward to see that there is a mapping for the appearance channel physics between NSI and the low scale NU frameworks~\cite{Blennow:2016jkn}: 
\be 
\label{eq:NUvsNSI}
2|\alpha_{\beta\gamma}|^2=|\epsilon^d_{\beta\gamma}|^2+|\epsilon^s_{\beta\gamma}|^2+2|\epsilon^d_{\beta\gamma}||\epsilon^s_{\beta\gamma}|\cos(\Phi^s_{\beta\gamma}-\Phi^d_{\beta\gamma})\, .
\ee

The NSI effects in disappearance channels show up already at linear order in $\epsilon$, $P_{\beta\beta}=1+\mathcal{O}(\epsilon)$. However, analogously to the NU case discussed above, the ND sensitivity via disappearance channels is limited by the normalization uncertainty. For this reason, we will focus only on the analysis of appearance channels. 

\section{Simulation details}
\label{sec:sim}

This section describes the main details used in our simulations for the DUNE experiment. In order to simulate the expected event rates at the DUNE ND, we use the same configuration as in the DUNE Technical Design Report (TDR), see Refs.~\cite{Abi:2020evt, Abi:2020qib}. All our calculations are performed using the GLoBES library~\cite{Huber:2004ka,Huber:2007ji}, together with the files provided by the collaboration as ancillary material with Ref.~\cite{Abi:2021arg}. Below we discuss the main relevant aspects of the simulation, while we refer the interested reader to Refs.~\cite{Abi:2021arg, Abi:2020qib, Abi:2020evt} for additional details. 

\subsection{Beam configuration and exposure}

\begin{figure}
\begin{center}
\includegraphics[scale=0.3]{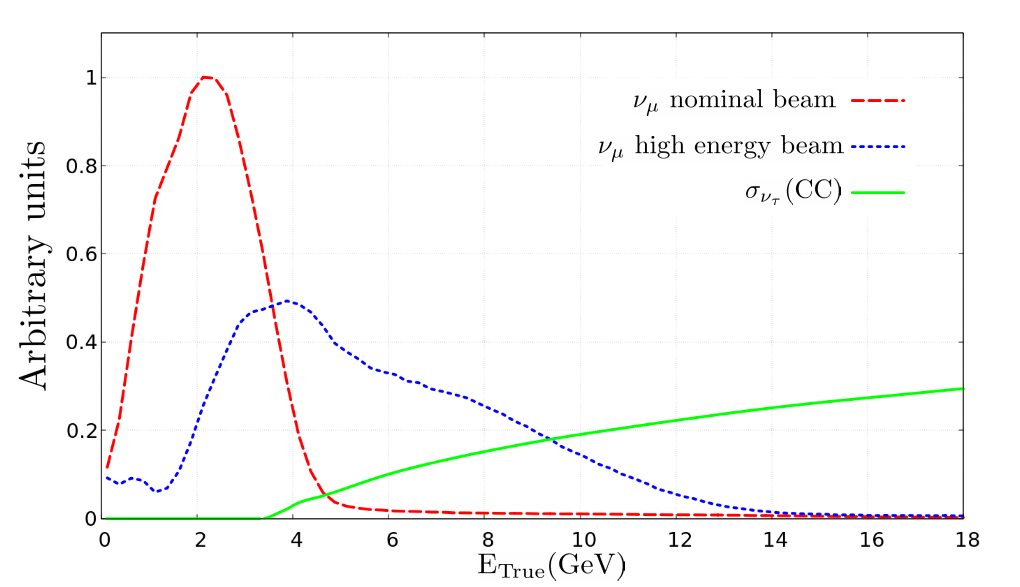}
\end{center}
\caption{\label{fig:flux} Comparison between the nominal $\nu_\mu$ flux and the HE flux as a function of the neutrino energy, in arbitrary units. Both curves are shown for neutrino mode only; the comparison is qualitatively similar for the antineutrino running mode fluxes. For comparison, the $\nu_\tau$ CC cross section is also shown.  }
\end{figure}

The main experimental configuration details are summarized in Tab.~\ref{tab:dune-config}. In particular, for the nominal running mode we consider equal exposure in neutrino and antineutrino running modes (3.5 years each, giving a total of 7 years of data taking), with a 1.2 MW beam.  
%
\begin{table}
\begin{center}
\begin{tabular}{ccccccc}
\hline
Beam configuration & Power & $E_{p}$ & PoT/yr & $t_\nu$ (yr) & $t_{\bar \nu}$ (yr) & $M_{\rm det}$ \\ \hline
Nominal & 1.2~MW & 120 GeV & $1.1\times10^{21}$ & 3.5 & 3.5 & 67.2 tons \\
High-Energy & 1.2~MW & 120 GeV & $1.1\times10^{21}$ & 3.5 & -- & 67.2 tons \\ \hline
\end{tabular}
\end{center}
\caption{\label{tab:dune-config} Exposure and beam configuration parameters used in the simulations. From left to right, the columns indicate beam power, proton energy, number of protons on target per year (PoT/yr), the running times in neutrino and antineutrino modes, and the fiducial mass of the detector.  }
\end{table}
%
The nominal neutrino flux for DUNE is shown in Fig.~\ref{fig:flux} as a function of the neutrino energy (dashed red line). For comparison, the solid green line in the same plot shows the $\nu_\tau$ charged-current (CC) cross section which, due to the large $\tau$ mass, does not kick in until the neutrino energy is at least 3~GeV. As shown in the figure the nominal neutrino flux peaks at much lower energies and, as a result, the ability to probe $\nu_\mu \to \nu_\tau$ conversion at the ND is going to be severely limited by statistics. For this reason, we have also considered adding 3.5 additional years of running time  (in neutrino mode) using the High-Energy (HE) beam configuration. This flux peaks at considerably higher energies as shown by the dotted blue line in Fig.~\ref{fig:flux} and therefore turns into a higher event rate for CC $\nu_\tau$ interactions.

\subsection{Simulation of the $\nu_\tau$-like event sample}

We use the $\nu_\tau$ cross section provided with the ancillary material of Ref.~\cite{Alion:2016uaj}, which corresponds to the $\nu_\tau$ CC cross section on Ar simulated with GENIE v2.8.4~\cite{Andreopoulos:2015wxa,Tena-Vidal:2021rpu}. Regarding the detection and reconstruction effects, the collaboration does not provide efficiencies or energy smearing matrices for a $\nu_\tau$ CC signal. We describe below our treatment of signal and backgrounds for the $\nu_\tau$-like event sample.  

Once a $\tau$ lepton has been produced at the detector, it will decay promptly. Given the neutrino energy is in the few GeV range, the distance traveled by the $\tau$ is not sufficiently long to allow for a successful particle identification at DUNE. However, its decay will still leave a visible signature in the detector: these events are characterized by either a hadronic shower or a charged lepton, plus missing energy carried away by the outgoing neutrino (or neutrinos, in the case of a leptonic decay). Given that the $\tau$ branching ratio into the hadronic decay channel is much larger than that of the leptonic channels (65\%, compared to a 17\% for each of the leptonic decay channels~\cite{Zyla:2020zbs}), in our analysis we include only those events where the $\tau$ decays hadronically.\footnote{We have explicitly checked that the inclusion of the electronic decay channel in our analysis (using the same efficiencies as in Ref~\cite{Abi:2021arg}) has no impact in the results.} For these events the main background comes from neutrino NC interactions, which would also yield a hadronic shower plus missing energy. Although the number of background events in this case is very significant, several cuts can be applied to increase the signal-to-background ratio, see Refs.~\cite{Conrad:2010mh,deGouvea:2019ozk, Machado:2020yxl}. Some of the most relevant signal/background discriminators include the energy of the most energetic pion, the number of pions produced in the shower, or the total visible energy of the event excluding the leading pion~\cite{Machado:2020yxl}. As a benchmark value we set the signal detection efficiency at 30\% based on Ref.~\cite{deGouvea:2019ozk}. 

The second relevant aspect of $\nu_\tau$ detection involves energy smearing. As outlined above, part of the incident neutrino energy will be carried away by the $\nu_\tau$ produced in the $\tau$ decay. Therefore, the visible energy will be significantly lower than the true incident neutrino energy. We implement this effect following Ref.~\cite{deGouvea:2019ozk}: for a given $E^{true}_\nu$, we assume that the observed energy distribution follows a Gaussian with mean value $0.45 E_\nu^{true}$ and width $0.25 E_\nu^{true}$. As for the background from NC events, we use the same migration matrices provided by the experimental collaboration for the $\nu_e$ CC channel. However, in this case we replace the background rejection efficiencies (which are provided for the $e$-like sample) by a constant 0.5\%, based on Ref.~\cite{deGouvea:2019ozk}. As a final comment we stress that the signal and background rejection efficiencies considered here are probably somewhat conservative and could be improved with the use of more sophisticated machine learning techniques, as demonstrated in Ref.~\cite{Machado:2020yxl} for the event sample where the $\tau$ decays into electrons. 

\subsection{Event rates}

\begin{figure}
\begin{center}
  \includegraphics[width=\textwidth]{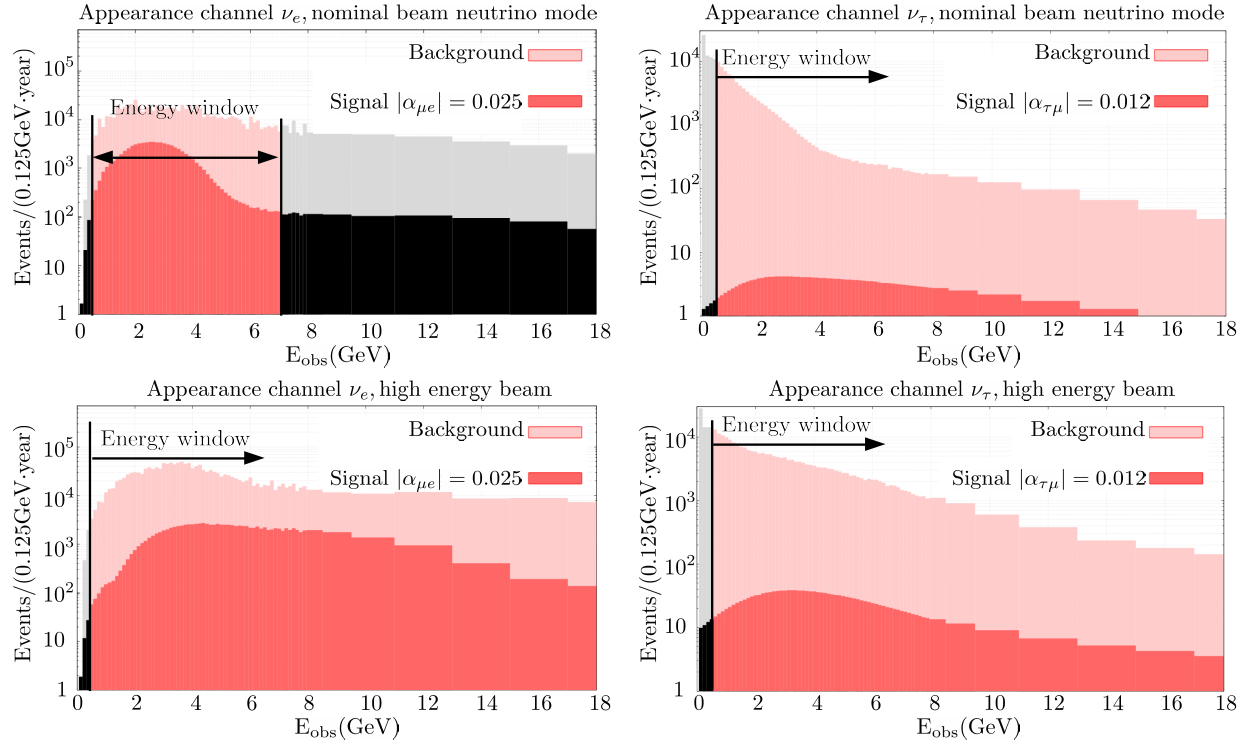}
\end{center}
\caption{\label{fig:histo} Event distributions as a function of the observed energy in the detector. Upper (lower) panels show the expected rates for the nominal (HE) beam configuration, for one year of data taking in neutrino mode. The light (dark) histograms correspond to the total background (signal) events for the NU case, with $\alpha_{\mu e}$ and $\alpha_{\tau \mu}$ as indicated by the labels. The resulting histograms for antineutrino mode are qualitatively similar to the upper panels and are not shown here. The colored bins (labeled as ``Energy window'') indicate the events included in the analysis, see text for details.}
\end{figure}

The expected event rates are shown in Fig.~\ref{fig:histo} for the $\nu_e$-like and $\nu_\tau$-like samples, as indicated by the labels. The total event rates for the different channels are summarized in Tab.~\ref{tab:events} in App.~\ref{app:events} for the nominal and HE beam modes. For illustration purposes, we show in Fig.~\ref{fig:histo} separately the background and signal contributions, for a value of the NU parameters which saturates the present bounds. As can be seen from the figure, even in the NU scenario the signal and background event rates present a very different dependence with the observed energy. In the $\nu_e$-like sample, this is because the leading contribution to the background comes from the intrinsic contamination of the beam, which has a very different shape compared to the leading $\nu_\mu$ component since it stems mostly from kaon instead of pion decays. In the case of the $\nu_\tau$-like sample, on the other hand, the fraction of energy carried away by the outgoing neutrino in NC scattering events and $\tau$ decays will be different, which translates into a different amount of migration towards lower values of the observed energy. 

Overall, the expected event rates shown in Fig.~\ref{fig:histo} illustrate the need to go beyond a naive implementation of systematic errors in terms of normalization uncertainties, and to include shape uncertainties as well.  The impact will be even larger in the sterile neutrino case (not shown here for conciseness), where the signal event rates would show an oscillatory pattern.

\subsection{Systematic uncertainties included in the fit}

At conventional long-baseline neutrino oscillation experiments the neutrino flux is produced from meson decays (predominantly pions and kaons) which, after being produced are focused into a decay pipe. Thus, the neutrino flux prediction cannot be computed analytically and relies on Monte Carlo simulations. Its associated systematic errors are largely dominated by hadron-production uncertainties, which can be as large as 10\%-15\% (see e.g. Ref.~\cite{Abi:2020evt} in the context of DUNE). Additional flux uncertainties come from beam focusing effects and can modify the expected shape of the neutrino flux, although these are typically kept at the few percent level. 

As shown in Fig.~\ref{fig:histo}, the nominal flux at DUNE peaks at approximately 2.5~GeV. At these energies, there is a similar contribution to the total neutrino event rates from quasi-elastic (QE), resonance production (RES) and deep-inelastic scattering (DIS) processes. While the DIS cross section can be accurately described starting from the neutrino interactions with the partons in the nucleons, the situation is very different for RES and QE processes, for which nuclear effects are relevant. For example it has been shown that the impact of two-particle-two-hole (2p2h) effects and final state interactions can lead to a significant bias in the neutrino energy reconstruction~\cite{Martini:2009uj,Martini:2010ex,Nieves:2011pp,Nieves:2011yp,Lalakulich:2012hs}, which in turn can have a large impact on the measurement of physics observables (see e.g. Ref.~\cite{Alvarez-Ruso:2017oui} for a recent review). Therefore, in what follows special attention has been paid to the role of shape uncertainties in our fits.

All simulations in this work have been performed using GLoBES v3.2.16~\cite{Huber:2004ka,Huber:2007ji}, which includes the same systematics implementation as in Ref.~\cite{Coloma:2012ji}. In order to account for shape uncertainties in our fit, a set of bin-to-bin uncorrelated nuisance parameters has been included for the most significant contributions to the total background. Additionally, for $\nu_\mu$-like events, the sensitivity is expected to be limited by the systematics on the $\nu_\mu \to \nu_\mu$ channel, and therefore shape uncertainties are also included for the signal in this case. An overall nuisance parameter (bin-to-bin correlated) is also included to allow for an overall change in normalization, for each signal and background channel separately, for all channels. A pull term is then added to the $\chi^2$ for each nuisance parameter, and the final $\chi^2$ is obtained after minimization over all nuisance parameters included in the fit. Table~\ref{tab:sys} summarizes the prior uncertainties included in our simulations, while a more detailed description of our $\chi^2$ implementation and the correlations implemented can be found in App.~\ref{app:chi2}.

\begin{table}
\begin{center}
\begin{tabular}{cc | cc | cc | cc}
\hline \hline
\multirow{2}{*}{Event sample} & \multirow{2}{*}{Contribution} & \multicolumn{2}{c}{Benchmark 1} & \multicolumn{2}{c}{Benchmark 2} & \multicolumn{2}{c}{Benchmark 3} \\ 
&  & $\sigma_{norm}$ & $\sigma_{shape}$ & $\sigma_{norm}$ & $\sigma_{shape}$  & $\sigma_{norm}$ & $\sigma_{shape}$ \\ \hline\hline
\multirow{4}{*}{$\nu_e$-like} & Signal              & 5\%    & --      & 5\%  & --   & 5\%   & --  \\
                                                    & Intrinsic cont.   & 10\% & --      & 10\% & 2\%  & 10\%  & 5\%  \\
                                                    & Flavor mis-ID  & 5\%   & --       & 5\%  & 2\%   & 5\%   & 5\%\\ 
                                                    & NC                   & 10\%  & --       & 10\% & 2\%  & 10\% & 5\%  \\ \hline 
\multirow{2}{*}{$\nu_\mu$-like}  & $\nu_\mu, \bar\nu_\mu$ CC (signal) & 10\% & -- & 10\% & 2\% & 10\% & 5\%  \\  
                                                          & NC                                                     & 10\% & -- & 10\% & 2\% & 10\% & 5\% \\ \hline 
\multirow{2}{*}{$\nu_\tau$-like} & Signal & 20\% & -- & 20\% & -- & 20\% & --  \\ 
                                                        & NC      & 10\% & -- & 10\% & 2\%  & 10\% & 5\%\\ \hline \hline
\end{tabular}
\end{center}
\caption{\label{tab:sys} Assumed prior uncertainties affecting the normalization and shape of the event rates in our simulations, for the three benchmark scenarios considered in this work. We assume the same uncertainties (although completely uncorrelated) for the HE and nominal beam configurations, as well as for the neutrino and antineutrino modes. The background contributions are separated into intrinsic beam contamination (intrinsic cont.), flavor mis-identification (flavor mis-ID) and neutral-current (NC) backgrounds. We have numerically checked that for the $\nu_e$-like and $\nu_\tau$-like events, including a shape uncertainty also for the signal has a completely negligible impact in the analysis, as expected. For additional details on the systematics implementation, see App.~\ref{app:chi2}.}
\end{table}
%
As can be seen from Tab.~\ref{tab:sys}, the priors assumed for the normalization uncertainties range between 5\% and 20\%, depending on the particular channel. Our reasoning for choosing these values is based on the fact that, although some of the predictions for the fluxes and cross sections rely on the prediction from Monte Carlo simulations (with uncertainties that can be as large as 10\% - 20\%), others may be reduced combining different measurements at the ND. For example, in the $\nu_\mu$-like sample we assume a 10\% normalization uncertainty, mainly driven by the large hadron production uncertainties affecting the  flux prediction from simulations. However, a measurement of the $\nu_\mu$ CC event sample at the ND can be used to normalize the flux for the contribution to the signal from $\nu_\mu \to \nu_e$ anomalous events, as well as the backgrounds from flavor mis-identification to this search. Thus, for these contributions we assume a smaller uncertainty, at the 5\% level. Similarly, while the neutrino NC cross section in the few GeV range is poorly known~\cite{Formaggio:2013kya}, a dedicated measurement of NC events at the DUNE ND can be used to reduce the uncertainties on the NC background for the $\nu_e$-like and $\nu_\tau$-like samples. In the case of shape uncertainties, driven by cross section and beam focusing effects, we consider two different benchmark values at the 2\% and 5\% as outlined in Tab.~\ref{tab:sys}, plus a third case where shape uncertainties are not included in the analysis.

\section{Results}
\label{sec:results}

This section summarizes the main results of our work. We will discuss separately the results obtained for the two frameworks under study, namely: NU of the leptonic mixing matrix coming from new physics at low scales, and sterile neutrinos participating in oscillations. For completeness, we will also provide the results for NSI in production and detection, which can be obtained from the mapping between NU and NSI as discussed in Sec.~\ref{sec:notation}.

\subsection{Non-unitarity}

In the non-unitarity case the observable impact on the event rates at the DUNE ND would be through a change in normalization. As a consequence, the sensitivity via disappearance channels to $\alpha_{ee}$ and $\alpha_{\mu\mu}$, see Eq.~(\ref{eq:PsterilesL0_3}), is expected to be limited by the size of the systematic errors affecting the normalization of the event rates for the $\nu_e$-like and $\nu_\mu$-like samples. For this reason, we do not consider these parameters here and focus instead on the determination of the off-diagonal parameters $\alpha_{\mu e}$ and $\alpha_{\tau\mu}$, which would lead to observable differences in the energy dependence of the total event rates, as shown by the histograms in Fig.~\ref{fig:histo}. 

\begin{figure}[ht!]
\begin{center}
  \includegraphics[width=0.97\textwidth]{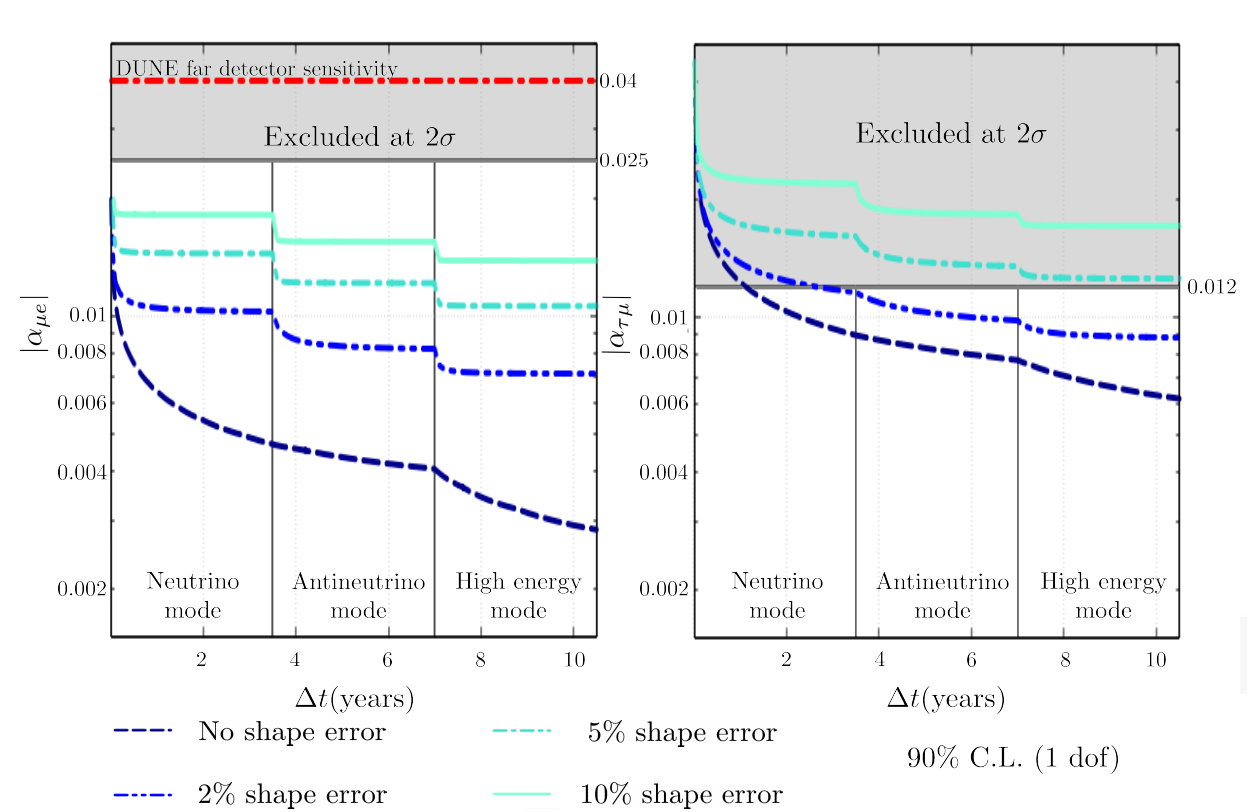} 
\end{center}
\caption{\label{fig:NU} Sensitivity to the off-diagonal NU parameters  $\alpha_{\mu e}$ (left panel) and $\alpha_{\tau\mu}$ (right panel). The lines show the sensitivity at 90\% CL for 1 degree of freedom (d.o.f.) as a function of the running time $\Delta t$ (in number of years), for different choices of the systematic uncertainties as indicated by the legend. The vertical lines indicate the changes between neutrino and antineutrino running modes (in the nominal beam scenario) as well as the change to the HE beam. For reference, the shaded gray areas show the region disfavored at 90\% CL by current constraints. For $\alpha_{\mu e}$, the dot-dashed red line indicates the expected sensitivity using the far detector data (taken from Ref.~\cite{Blennow:2016jkn}), while for $\alpha_{\tau\mu}$ the expected sensitivity would be worse than the range shown in the figure and is therefore not shown.}
\end{figure}

Our results  are shown in Fig.~\ref{fig:NU} for $\alpha_{\mu e}$ (left panel) and $\alpha_{\tau\mu}$ (right panel), respectively. The different lines show the results for different choices of systematic uncertainties as indicated by the labels, as a function of running time. Two main features can be seen from both panels in the figure. First, the results depend severely on the choice of systematic errors (as expected). Second, if shape uncertainties are included in the analysis the results are completely dominated by systematics and the sensitivity does not improve significantly by increasing the number of years of data taking in each mode. There is a certain improvement in sensitivity due to the changes between neutrino and antineutrino modes, as well as from the addition of extra data in the HE mode: this is due to new information coming into the fit from a different dependence of the signal-to-background ratio with the energy in the different running modes. However, after an initial (sharp) increase in sensitivity the experiment enters again a systematics-dominated regime and the results stop improving with new data. 

Our results show an expected improvement with respect to current bounds for $\alpha_{\mu e}$, as shown in the right panel. In the case of $\alpha_{\tau\mu}$, however, the outcome will depend on the final level of systematic uncertainties: even in the case in which a 2\% shape uncertainty is assumed the improvement over current bounds on this parameter is marginal. This is so because of the much larger background levels expected for the $\nu_\tau$ samples with respect to the signal (as shown in Fig.~\ref{fig:histo}).

\subsection{Sterile neutrinos}

Before discussing our results, let us point out that the determination of the sensitivity to sterile neutrinos from oscillations is non-trivial from the statistical point of view. In particular, the results obtained under the assumption that Wilks' theorem~\cite{Wilks:1938dza} applies may differ from the ones obtained via Monte Carlo simulation~\cite{Agostini:2019jup,Almazan:2020drb,Giunti:2020uhv, Coloma:2020ajw}. The sensitivity contours in this work have however been computed under the assumption that Wilks' theorem applies, in order to ease the comparison with previous literature. 

In order to discuss the sensitivity to sterile neutrino oscillations it is important to keep in mind that at the DUNE ND the event rates would be sensitive to several oscillation probabilities: $P_{ee}$, $P_{\mu\mu}$, $P_{\mu e}$ and $P_{\mu\tau}$. Each of them will be sensitive to a different set of mixing matrix elements as described in Sec.~\ref{sec:notation}, see Eq.~\eqref{eq:sterile-eff-angles}. In the $3+1$ framework it is also common to parameterize the full mixing matrix as $\mathcal{U} = R_{34} S_{24} S_{14} R_{23} S_{13} R_{12}$, where $R_{ij}$ is a rotation matrix in the $ij$ sector with mixing angle $\theta_{ij}$, while $S_{14}$ are complex rotation matrices which also include a CP phase $\delta_{ij}$. The new rotation angles driving the mixing between the mostly sterile mass eigenstate and the active neutrinos can be trivially mapped to the triangular parameterization, and also be written in terms of mixing matrix elements as:
\bea
|\mathcal{U}_{e4}|^2&=&2\alpha_{ee}=s_{14}^2,\nonumber\\
|\mathcal{U}_{\mu 4}|^2&=&2\alpha_{\mu\mu}=s_{24}^2+\mathcal{O}(s_{24}^2s_{14}^2)\nonumber\\
|\mathcal{U}_{\tau 4}|^2&=&2\alpha_{\tau\tau}=s_{34}^2+\mathcal{O}(s_{24}^2s_{34}^2)+\mathcal{O}(s_{34}^2s_{14}^2)\nonumber.
\eea
%

\begin{figure}[ht!]
\begin{center}
  \includegraphics[width=0.65\textwidth]{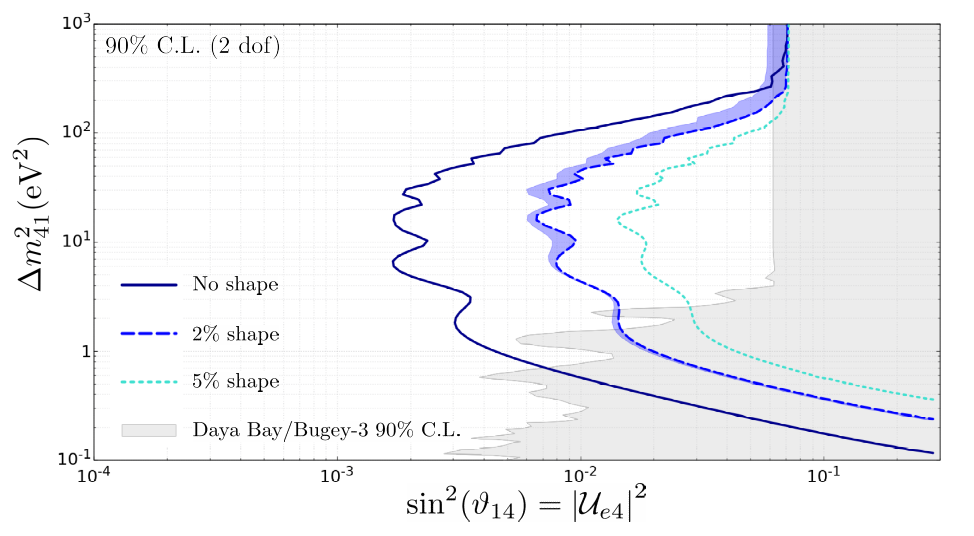}
  \includegraphics[width=0.65\textwidth]{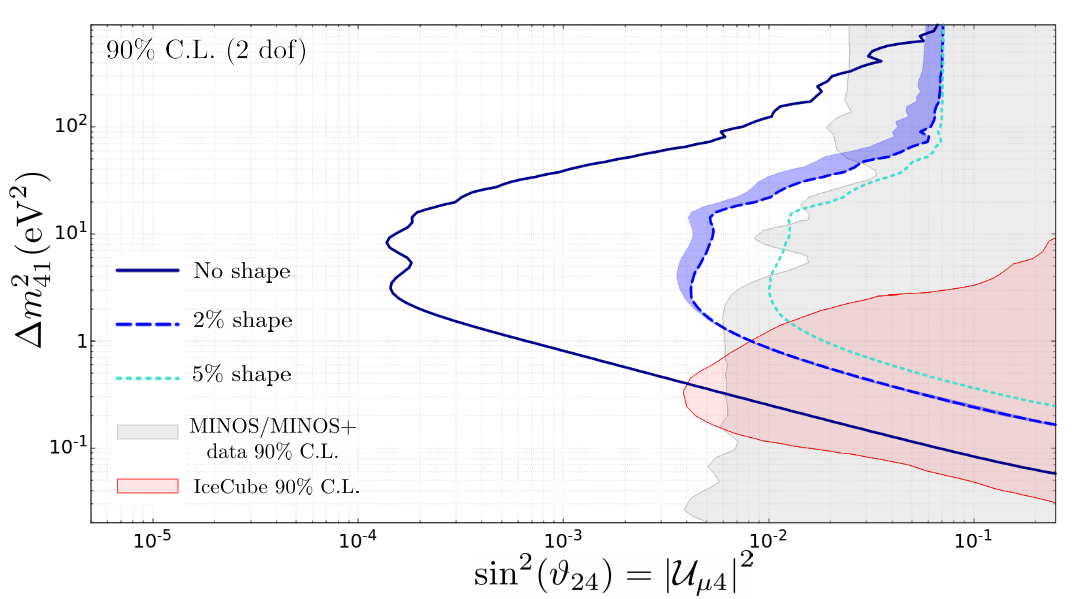}
  \includegraphics[width=0.65\textwidth]{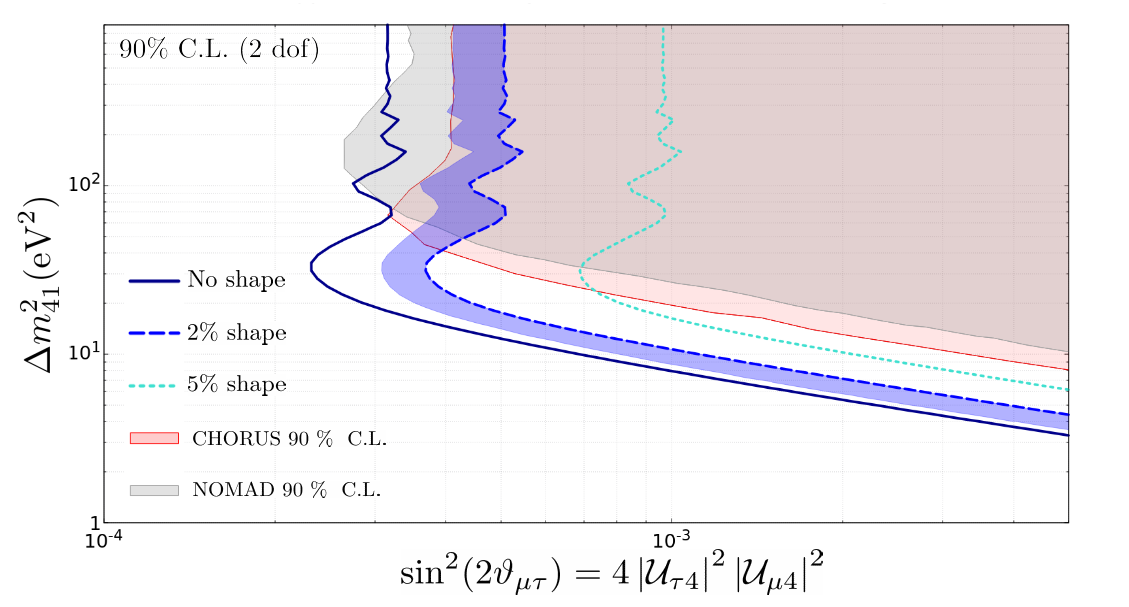}
\end{center}
\caption{\label{fig:sterile} Expected sensitivity to the sterile neutrino scenario, for oscillations in the $P_{ee}$, $P_{\mu\mu}$ and $P_{\mu\tau}$  channels for the top, middle and lower panels, respectively. The shaded regions show current constraints from other experiments~\cite{Astier:2001yj,Tsenov:2009jca,TheIceCube:2016oqi,Adamson:2020jvo}, while the colored lines indicate the sensitivity for the DUNE ND, for different choices of systematic uncertainties as indicated by the legend. In each case, the region to the right of the colored lines would be disfavored at 90\% CL (2 d.o.f.). Finally, the colored blue bands indicate the increase in sensitivity due to the addition of 3.5 years of data taken in the HE mode.  }
\end{figure}

In Fig.~\ref{fig:sterile} we present the results obtained for the sensitivity to a sterile neutrino inducing non-standard $P_{ee}$, $P_{\mu\mu}$ and $P_{\mu\tau}$ probabilities at the DUNE ND. In each case, the sensitivity comes mainly from the measurement of the $\nu_e$-like, $\nu_\mu$-like and $\nu_\tau$-like samples separately. Therefore, no combination between different samples is performed for any of the results shown in this figure. For comparison, we compare our results to the present dominant constraints from Daya Bay/Bugey-3~\cite{Adamson:2020jvo}, NOMAD~\cite{Astier:2001yj}, CHORUS~\cite{Tsenov:2009jca}, MINOS/MINOS+~\cite{Adamson:2020jvo}, and Icecube~\cite{TheIceCube:2016oqi}. Note that Eq.~\eqref{eq:Psteriles} implies that the oscillation probability is invariant under the change $|\mathcal{U}_{\alpha 4}|^2 \to (1-|\mathcal{U}_{\alpha 4}|^2)$. Therefore, a null result in the disappearance channels can be accommodated not only for small values of  $\mathcal{U}_{\alpha 4}$ (as expected) but for large values as well. Very large values of the mixing matrix elements in the fourth column of the PMNS are however excluded by present neutrino oscillation data~\cite{Parke:2015goa}. Thus, in order to be consistent with current bounds we only consider values of the mixing matrix elements such that $|\mathcal{U}_{\mu (e) 4}|^2 < 0.5$. 

As in the case for the NU scenario, in the sterile neutrino case we see again a large dependence of the results with the implementation of systematic uncertainties. The effect is dramatic on the sensitivity to $P_{\mu\mu}$ oscillations, as shown in the middle panel of Fig.~\ref{fig:sterile}, where shape uncertainties at the level of 2\% lead to a decrease in sensitivity of over an order of magnitude with respect to the scenario where only normalization uncertainties are considered. In this case, only a mild improvement over present constraints from MINOS could be expected at DUNE. The reason is that, in this channel, the sensitivity is almost entirely driven by the observation of a small oscillatory pattern on top of the expected (large) number of $\nu_\mu$ CC events (see Tab.~\ref{tab:events} in App.~\ref{app:events}), which can be easily hidden by the inclusion of shape uncertainties. The effect is similar (although a bit milder) for the $P_{ee}$ case shown in the top panel of Fig.~\ref{fig:sterile}; however, current constraints in the $\Delta m_{41}^2 \sim 10~\rm{eV}^2$ region are not as strong as in the $P_{\mu\mu}$ case and, therefore, there is still room for a considerable improvement (up to one order of magnitude) if shape uncertainties are at the 2\% level. Finally, we also present our results for the $P_{\mu\tau}$ channel in the lower panel of Fig.~\ref{fig:sterile}. In this case, the sensitivity is severely limited by the reduced statistics; however, for shape uncertainties at the level of 2\%, there is room for a factor $\sim 5$ improvement over current bounds from CHORUS~\cite{Tsenov:2009jca} and NOMAD~\cite{Astier:2001yj} for mass-squared splittings $\Delta m_{41}^2 \lesssim 50~\rm{eV}^2$, specially if HE data is included in the fit (indicated by the shaded blue band).

Figure~\ref{fig:sterile} also shows how the average-out regime (in which the sensitivity becomes independent of $\Delta m_{41}^2$) is achieved for $\Delta m_{41}^2 \gtrsim \mathcal{O}(100)~\rm{eV}^2$. Notice that, for the $P_{\mu\mu}$ and $P_{ee}$ channels, the sensitivity in this region is basically limited by the normalization uncertainty, see Eq.~\eqref{eq:PsterilesL0_3}. Since we are using similar normalization uncertainties for both channels  (see Tab.~\ref{tab:sys}), a similar limit is obtained for  $|\mathcal{U}_{e 4}|^2$ and $|\mathcal{U}_{\mu 4}|^2$ in this regime. For the  $P_{\mu\tau}$ channel, we also observe that in the averaged-out region the low-scale NU limit shown in Fig.~\ref{fig:NU} is recovered, as expected from  Eq.~(\ref{eq:PsterilesL0_2}).

\begin{figure}[ht!]
\begin{center}
  \includegraphics[width=0.95\textwidth]{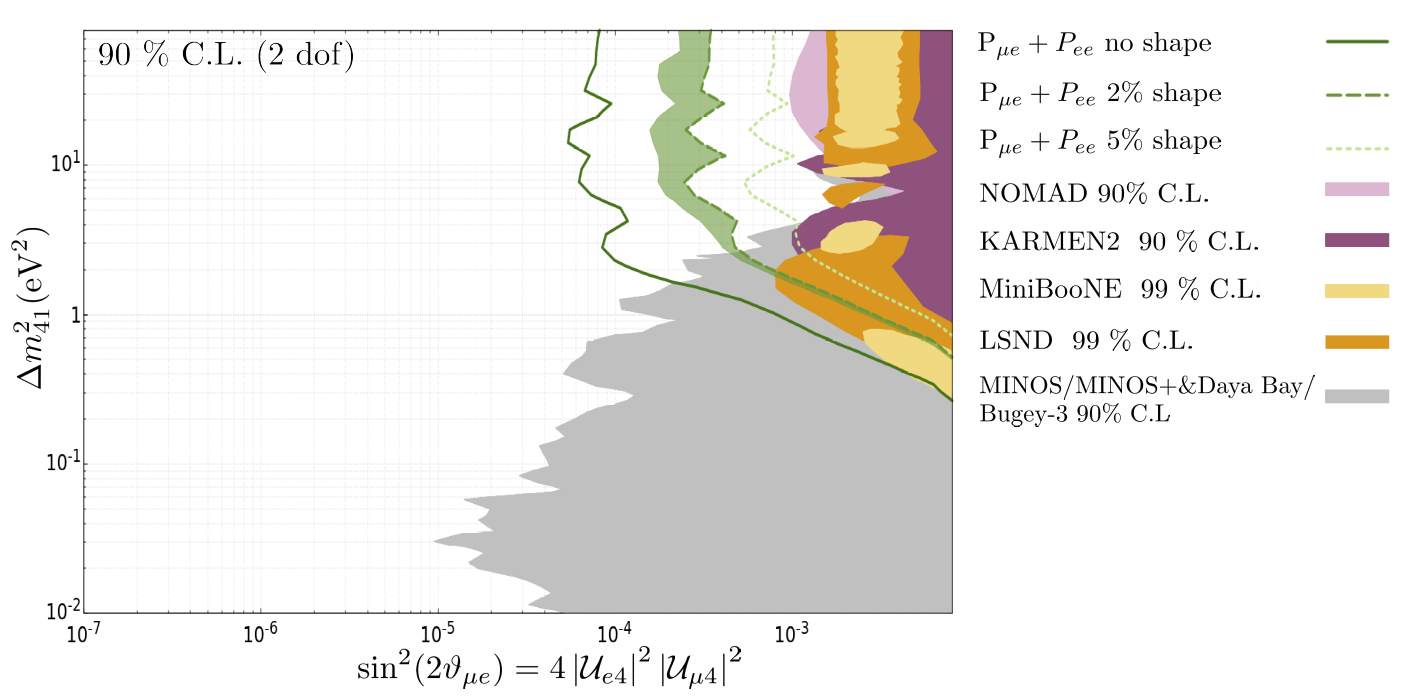}
  \includegraphics[width=0.95\textwidth]{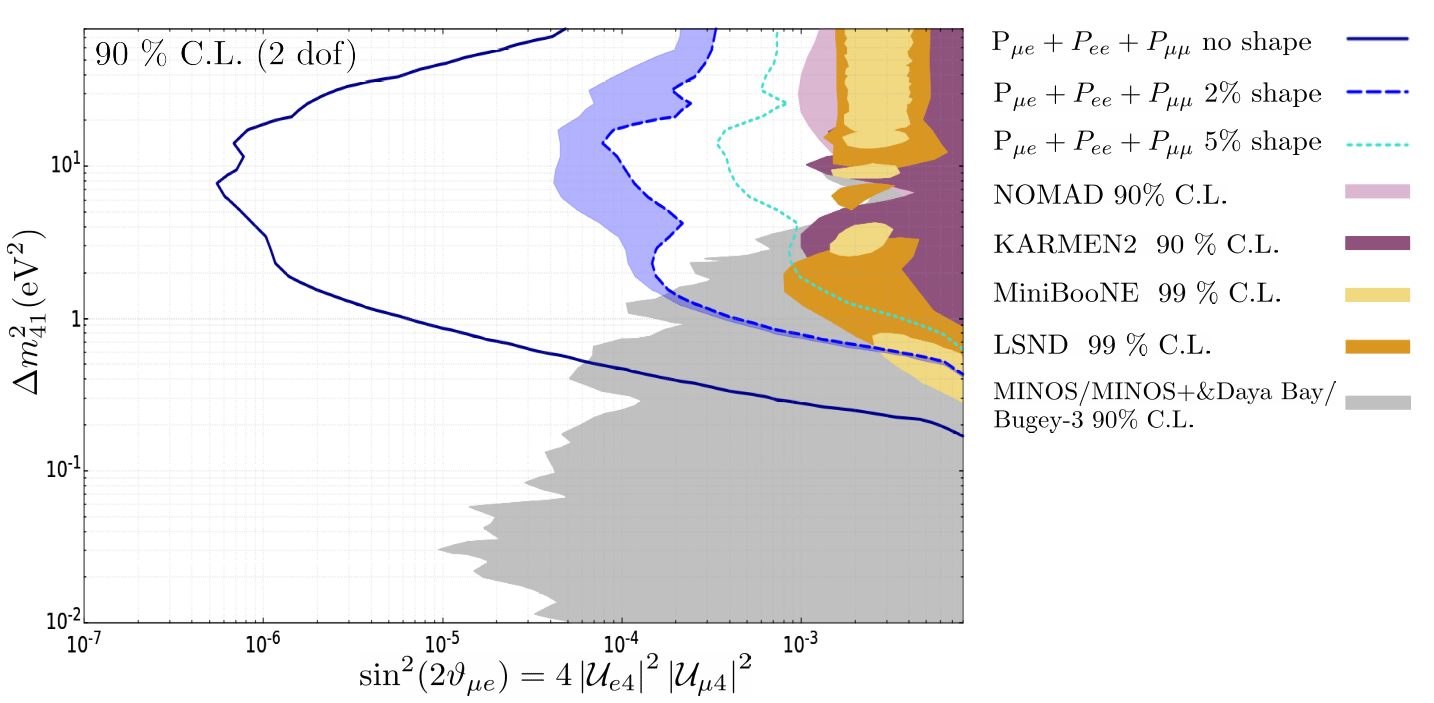}
\end{center}
\caption{\label{fig:sterile-Pmue} Expected sensitivity to the sterile neutrino scenario, for oscillations in the $P_{\mu e}$ channel. Results in the upper panel only include information from the $\nu_e$-like sample, while in the lower panel we have added the information from the $\nu_\mu$-like sample as well. The shaded pink, purple and gray areas show current constraints from other experiments at 90\% CL~\cite{Armbruster:2002mp,Astier:2003gs,Adamson:2020jvo}, while the shaded yellow/orange regions are favored at the 99\% CL by the LSND~\cite{Aguilar:2001ty} and MiniBooNE~\cite{Aguilar-Arevalo:2020nvw} anomalies. The colored lines indicate the sensitivity for the DUNE ND, for different choices of systematic uncertainties as indicated by the legend. In each case, the region to the right of the lines would be disfavored at 90\% CL (2 d.o.f.). Finally, the shaded band to the left of the dashed lines indicate the increase in sensitivity due to the addition of  3.5 years of data taken in the HE mode.}
\end{figure}

Figure~\ref{fig:sterile-Pmue}, on the other hand, shows our results for oscillations in the $P_{\mu e}$ channel. Again, for comparison the shaded pink, purple and gray areas show current dominant constraints from previous experiments, namely, NOMAD~\cite{Astier:2003gs}, KARMEN~\cite{Armbruster:2002mp} and the combination of MINOS/MINOS+ and Daya Bay/Bugey-3 data~\cite{Adamson:2020jvo}. Note that while NOMAD and KARMEN directly probed the $P_{\mu e}$ channel, the combination of MINOS/MINOS+ and Daya Bay/Bugey-3 provides an indirect constraint from $P_{ee}$ and $P_{\mu\mu}$ disappearance channels. For reference, the shaded yellow/orange regions indicate the fraction of parameter space that is favored at 99\% CL by the LSND~\cite{Aguilar:2001ty} and MiniBooNE~\cite{Aguilar-Arevalo:2020nvw} anomalies.

In principle the sensitivity to the $P_{\mu e}$ channel comes from the measurement of the $\nu_e$-like sample, which is directly sensitive to transitions between muon and electron neutrinos. However, at DUNE the combination with data from the $\nu_\mu$-like sample would greatly enhance the sensitivity since it would allow for a simultaneous constraint on $\mid\!\mathcal{U}_{e 4}\!\mid^2$, $\mid\!\mathcal{U}_{\mu 4}\!\mid^2$, and $\mid\!\mathcal{U}_{e 4}\!\mid^2\mid\!\mathcal{U}_{\mu 4}\!\mid^2$. This can be seen from the comparison between the top and lower panels in Fig.~\ref{fig:sterile-Pmue}: while in the top panel the analysis is performed using only information from the $\nu_e$-like sample (which is sensitive to $P_{\mu e}$ and $P_{ee}$), in the lower panel we also include the information on $P_{\mu\mu}$ from the $\nu_\mu$-like sample. 

In fact, the combination of data from different channels for a sterile neutrino search was already performed for the DUNE experiment in Ref.~\cite{Abi:2020kei}, leading to an impressive sensitivity to sterile neutrinos which reached values of the effective mixing angle of $\sin^22\theta_{\mu e} \lesssim 10^{-6}$ in the region where $\Delta m_{41}^2 \sim \mathcal{O}(1-10)~\rm{eV}^2$. However, in Ref.~\cite{Abi:2020kei} only normalization uncertainties were included in the analysis. Interestingly, even though our implementation of normalization uncertainties is very different than in Ref.~\cite{Abi:2020kei}  we recover a similar result here, shown by the solid blue line in the lower panel of Fig.~\ref{fig:sterile-Pmue}. This already indicates that normalization uncertainties are not playing a relevant role in the fit, which is expected since the effect from a sterile neutrino would be detected by looking at spectral distortions in the event rates. Indeed, once shape uncertainties are included at the 2\% level (shown by the dashed blue lines in the lower panel) the corresponding result is worsened by two orders of magnitude with respect to the case where only normalization uncertainties are included. Nevertheless, we find that even in this case the DUNE ND data should be able to probe most of the regions favored by the LSND and MiniBooNE experiments at more than 90\% CL, with the exception of the region at very low $\Delta m^2_{41}$ (which is in any case already disfavored by the combination of MINOS/MINOS+ and Daya Bay/Bugey-3 data). Furthermore, notice that DUNE has the additional advantage of being able to consistently probe the sterile neutrino hypothesis with the same experimental setup using different and complementary neutrino oscillation channels: $P_{ee}$, $P_{\mu\mu}$ and $P_{\mu e}$.

\subsection{Non-Standard Interactions in production and detection}

\begin{figure}[ht!]
\begin{center}
  \includegraphics[width=0.47\textwidth]{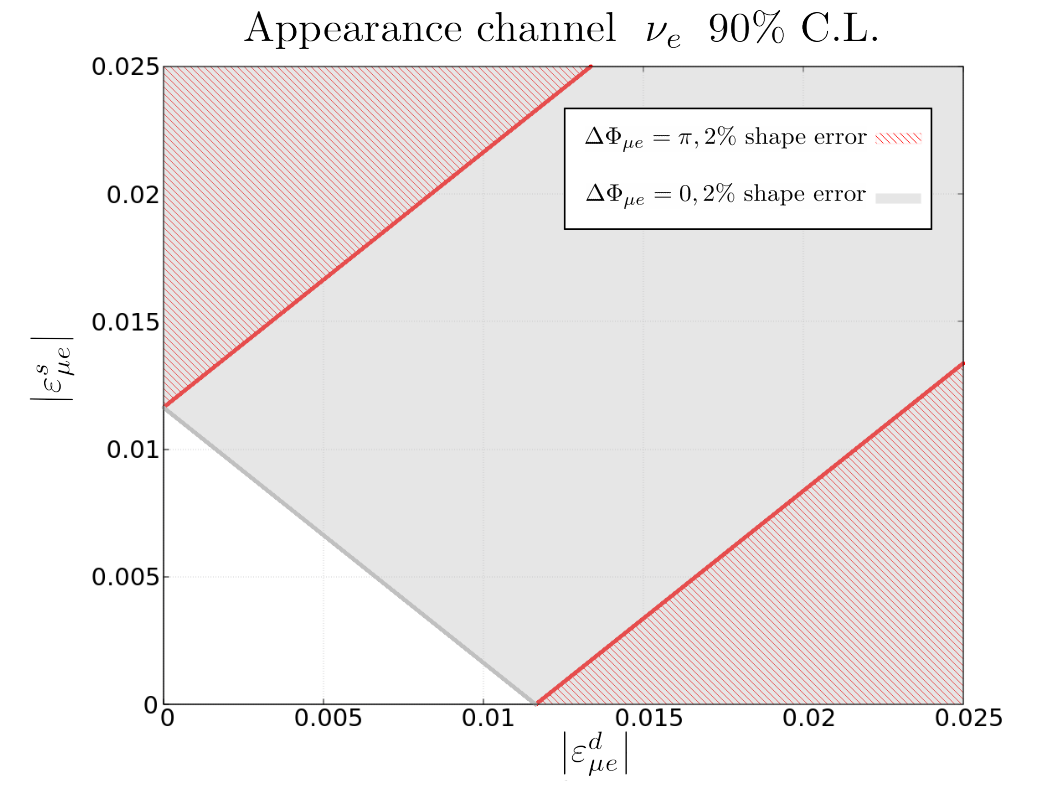}
  \includegraphics[width=0.48\textwidth]{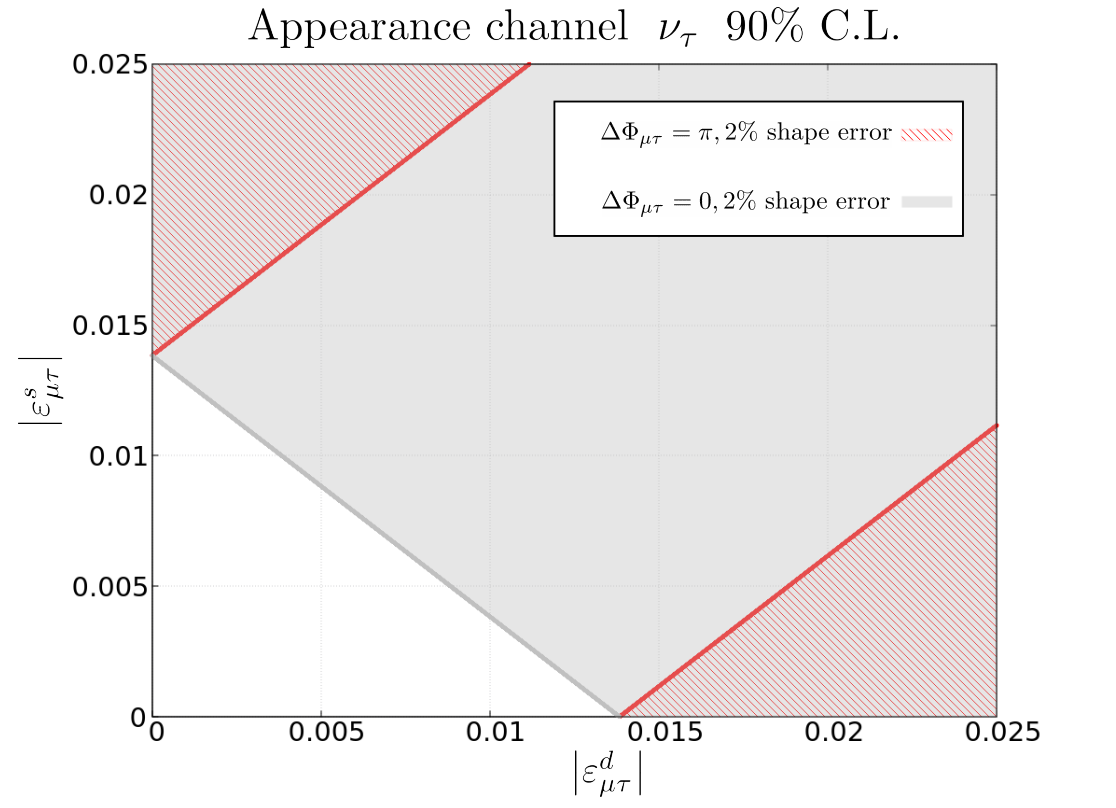}
\end{center}
\caption{\label{fig:NSI} Sensitivity to NSI in production and detection. Results are shown for effects on the $P_{\mu e}$ (left panel) and $P_{\mu\tau}$ (right panel). In both panels the sensitivity is shown for the two limiting cases $\Delta \Phi = \pi, 0$, which lead to a destructive and constructive interference between production and detection NSI respectively  (see Eq.~\eqref{eq:prob-NSI}). }
\end{figure}

Finally, for completeness we also include an explicit recasting of our bounds from low scale NU to the NSI formalism in production and detection, following the prescription from Eq.~(\ref{eq:NUvsNSI}). Our results are shown in Fig.~\ref{fig:NSI}, for NSI parameters affecting $P_{\mu e}$ (left panel) and $P_{\mu\tau}$ (right panel). 

As can be seen from Eq.~\eqref{eq:prob-NSI}, the oscillation probability in this case depends on the production and detection NSI including their possible CP-violating phases, which come in through an interference term that is proportional to $\cos(\Phi^s_{\beta\gamma} - \Phi^d_{\beta\gamma}) \equiv \cos\Delta\Phi_{\beta\gamma}$. For this reason we show in both panels in Fig.~\ref{fig:NSI} the sensitivity to production and detection NSI in two limiting cases, depending on the value of $\Delta \Phi$.

Our results for NSI can be compared to those obtained in Ref.~\cite{Giarnetti:2020bmf}. Regarding the sensitivity in the disappearance channels, 
we find that the results are essentially limited by the normalization uncertainties (as expected from naive theoretical arguments) and are therefore not shown here. On the other hand, for the appearance channels we have checked that, when no shape uncertainties are considered, our results for the $P_{\mu e}$ and $P_{\mu\tau}$ channels agree relatively well with those obtained in Ref.~\cite{Giarnetti:2020bmf}. 
For the $P_{\mu\tau}$ channel we have checked that including in our analysis the additional events from the $\tau$ decay channel into electrons has a negligible impact on our results. 

\section{Summary and conclusions}
\label{sec:conclusions}

Oscillation experiments using conventional neutrino beams rely on the use of a near detector (ND), located at typical distances of $\mathcal{O}(500)$~m from the target station, to measure the unoscillated neutrino event rates. In this work we have studied the potential of the ND at the future DUNE experiment to test new physics affecting oscillations in three scenarios: effects coming from non-unitarity (NU) of the leptonic mixing matrix due to new states at low scales, light sterile neutrino oscillations, and Non-Standard Interactions (NSI) affecting neutrino production and detection processes. While sterile neutrinos would induce a new oscillation at the ND, NU and NSI would lead to a so-called zero-distance effect on the oscillation probabilities. 

Using the ND to search for new physics is, however, technically challenging due to the large systematic uncertainties associated to the neutrino flux and cross sections. While most works in previous literature include only normalization uncertainties, for the scenarios considered in this work shape uncertainties are most important. Our results show that shape uncertainties at the 2\% level can lead to a reduction in the  sensitivity to sterile neutrino oscillations in the $P_{\mu e}$ channel of two orders of magnitude, and around one order of magnitude in the case of oscillations in the $P_{ee}$ and $P_{\mu\mu}$ channels. Their impact is also relevant for the NU scenario, leading to a decrease in sensitivity of about a factor of 2 (3) in the case of $\alpha_{\mu e}$ for shape uncertainties at the 2\% (5\%) level. Our results qualitatively agree with those presented in Ref.~\cite{Miranda:2018yym}, where the sensitivity to both scenarios (including shape uncertainties) via the $P_{ee}$, $P_{\mu\mu}$ and $P_{\mu e}$ channels was studied (albeit for different ND configurations).

We have also studied the sensitivity to new physics leading to $\nu_\mu \to \nu_\tau$ transitions. In this case we find that the sensitivity is limited by the low signal-to-background ratio. This stems from the fact that the charged-current $\nu_\tau$ cross section starts to grow for neutrino energies $E_\nu \gtrsim 3.5~\rm{GeV}$, above the region where the nominal DUNE neutrino flux reaches its maximum. Our results show that this is specially relevant for the NU scenario, for which the DUNE ND will only be able to improve marginally over current constraints if shape uncertainties can be kept at the 2\% level or below. 

Finally, we have studied the impact on the results of additional data taken with the neutrino beam in the so-called High-Energy (HE) configuration. Being much more energetic, this leads to a slightly better sensitivity to new physics effects in the $P_{\mu\tau}$ channel. However, for the rest of oscillation channels considered in this work, the impact of the HE data is always rather minor since the sensitivities are mostly limited by systematic errors instead of statistics. 

In summary, our results show that the potential of the DUNE ND to constrain new physics scenarios affecting oscillations at short baselines is severely limited by shape uncertainties, an effect that is typically overlooked in the literature. This stresses the importance of a joint experimental and theoretical effort to improve our understanding of neutrino nucleus cross sections, as well as hadron production uncertainties and beam focusing effects. Nevertheless, even with our more conservative and realistic implementation of systematic uncertainties, our results indicate that an improvement over current bounds is generally expected.

\acknowledgments{The authors warmly thank Justo Mart\'in Albo, Laura Molina Bueno, Holger Schulz, Jessica Turner and Elizabeth Worcester for useful discussions. SR thanks specially Enrique Fern\'andez-Mart\'inez, Mattias Blennow and Alexandre Sousa for useful discussions. This project has received funding/support from the European Union’s Horizon 2020 research and innovation program under the Marie Skłodowska-Curie grant agreement No 860881-HIDDeN. The authors acknowledge use of the HPC facilities at the IFIC (SOM cluster), and support of the Spanish Agencia Estatal de Investigación through the grant “IFT Centro de Excelencia Severo Ochoa SEV-2016-0597”. JLP and SU acknowledge support from Generalitat Valenciana through the plan GenT program (CIDEGENT/2018/019) and from the Spanish MINECO under Grant FPA2017-85985-P. The work of PC is supported by the Spanish MICINN through the ``Ram\'on y Cajal'' program under grant RYC2018-024240-I. SR acknowledges support of the Spanish Agencia Estatal de Investigaci\'on and the EU ``Fondo Europeo de Desarrollo Regional'' (FEDER) through the projects PID2019-108892RB-I00/AEI/10.13039/501100011033.}

\appendix

\section{Total event rates}
\label{app:events}

In this work the event rates are computed following Ref.~\cite{Abi:2021arg}. While we refer the interested reader to the mentioned references for details, here we provide a summary of the total event rates considered in our analysis, which are useful to understand the final results from our simulations. 

The total event rates are summarized in Tab.~\ref{tab:events} for the nominal and HE beam configurations. They are given separately for the contributions from: intrinsic contamination ($\nu_e$ and $\bar\nu_e$) of the beam; $\nu_\mu$ and $\bar\nu_\mu$ CC events mis-identified as $\nu_e$ or $\bar\nu_e$ events; and NC events mis-identified as CC events (in all cases, adding up the neutrino and antineutrino contributions to the event rates). The total number of $\nu_\mu$ and $\bar\nu_\mu$ CC events are also provided, for $P_{\mu\mu} =1$. The number of $\nu_e$ and $\bar\nu_e$ CC events can be estimated from this multiplying by the corresponding $P_{\mu e}$ probability (up to small differencies in detection efficiencies for the two flavors). The total number of $\nu_\tau$ and $\bar\nu_\tau$ events is also provided for $P_{\mu\tau}=1$, and can be trivially rescaled down for a given value of the NU parameters, see Eqs.~(\ref{eq:PsterilesL0_1}) and (\ref{eq:PsterilesL0_2}). All event rates are provided after efficiencies and smearing effects are taken into account. The numbers provided correspond to the total rates within a given energy window above 0.5 GeV and below the maximum observed energy given in the last column of the table. The binning is taken as in Ref.~\cite{Abi:2021arg}.

\begin{table}[htb!]
\begin{center}
\begin{tabular}{c|cccc}
\hline \hline
Running mode & Sample & Contribution & Event rates ($\times 10^5$) & $E_{\rm obs}^{\rm max}$ (GeV)\\ \hline\hline
\multirow{7}{*}{$\nu$ mode (nominal) } & \multirow{3}{*}{$\nu_e$-like} & Intrinsic cont. & 20.18  
&    \\
 &  & Flavor mis-ID & 4.61 
& 7.125  \\ 
 &  & NC & 6.77 
&   \\ \cline{2-5}
 & \multirow{2}{*}{$\nu_\mu$-like} & $\nu_\mu, \bar\nu_\mu$ CC ($P_{\mu\mu}=1$) & 2,235.72 
& \multirow{2}{*}{7.125}  \\  
 &  & NC & 17.35 
&  \\ \cline{2-5} 
 & \multirow{2}{*}{$\nu_\tau$-like} & $ \nu_\tau, \bar\nu_\tau$ CC ($P_{\mu\tau}=1$) & 39.33
& \multirow{2}{*}{18} \\ 
 &  & NC &  3.23 
&    \\ \hline 
\multirow{7}{*}{$\bar\nu$ mode (nominal)} & \multirow{3}{*}{$\bar\nu_e$-like} & Intrinsic cont. & 11.18  
&  \\
 &  & Flavor mis-ID & 1.07 
& 7.125 \\ 
 &  & NC & 3.89 
&  \\ \cline{2-5} 
 & \multirow{2}{*}{$\bar\nu_\mu$-like} & $\nu_\mu, \bar\nu_\mu$ CC  ($P_{\mu\mu}=1$) & 1,013.42 
&  \multirow{2}{*}{7.125} \\ 
 &  & NC & 9.76 
&  \\ \cline{2-5} 
 & \multirow{2}{*}{$\bar\nu_\tau$-like} & $ \nu_\tau, \bar\nu_\tau$ CC ($P_{\mu\tau}=1$) & 27.75
& \multirow{2}{*}{18} \\ 
 &  & NC & 1.80 
&   \\ \hline \hline
\multirow{7}{*}{$\nu$ mode (HE)} & \multirow{3}{*}{$\nu_e$-like} & Intrinsic cont. & 38.10 
&   \\
 &  & Flavor mis-ID & 12.98 
& 18 \\ 
 &  & NC & 30.51 
&  \\ \cline{2-5}
 & \multirow{2}{*}{$\nu_\mu$-like} & $\nu_\mu, \bar\nu_\mu$ CC ($P_{\mu\mu}=1$) & 5,784.30  
& \multirow{2}{*}{18} \\  
 &  & NC & 72.15 
&  \\ \cline{2-5} 
 & \multirow{2}{*}{$\nu_\tau$-like} & $ \nu_\tau, \bar\nu_\tau$ CC ($P_{\mu\tau}=1$) & 259.67
& \multirow{2}{*}{18} \\ 
 &  & NC & 9.42 
& \\ \hline \hline

\end{tabular}
\end{center}
\caption{\label{tab:events} Expected total number of events for the different contributions to the event samples simulated in our analysis, for the nominal and high-energy (HE) beam configurations. Note that for the nominal beam we consider 3.5 years of data taking in neutrino and in antineutrino modes, while for the HE configuration we only consider 3.5 years in neutrino mode, see Tab.~\ref{tab:dune-config}. The background contributions are separated into intrinsic beam contamination (intrinsic cont.), flavor mis-identification (flavor mis-ID) and neutral-current (NC) backgrounds, adding the contributions from neutrinos and antineutrinos. We also provide the expected number of charged-current (CC) $\nu_\mu$ and $\bar\nu_\mu$ events for $P_{\mu\mu}=1$, as well as the $\nu_\tau$ and $\bar\nu_\tau$ CC events for $P_{\mu\tau}=1$. The latter can be trivially rescaled to a given value of $\alpha$, see Eqs.~(\ref{eq:PsterilesL0_1}) and (\ref{eq:PsterilesL0_2}). In all cases, detection efficiencies and smearing effects have already been accounted for. Finally, the last column indicates the maximum energy considered for each of the samples (corresponding to observed energy). See text for additional details. }
\end{table}

\section{Details on the $\chi^2$ implementation used}
\label{app:chi2}

In order to account for shape uncertainties in our fit, a set of bin-to-bin uncorrelated nuisance parameters has been included in the analysis. Given the large number of bins involved in the analysis and, in order to keep the total number of nuisance parameters as low as possible, these are applied to the final background event rates (adding up all the background contributions listed in Tab.~\ref{tab:events}) and to the overall signal event rates as well (adding up the neutrino and antineutrino contributions to the signal). The shape nuisance parameters in the $i$-th energy bin are denoted as  $\xi^{\rm sig}_i$ or $\xi^{\rm bg}_i$ for the signal and background, respectively. An independent set of nuisance parameters (bin-to-bin correlated) is also included to allow for an overall change in normalization, in this case for each signal and background channel separately ($\zeta_c$, for a given contribution $c\equiv s,b$ to the signal or background rates). Thus, for each sample a Poissonian $\chi^2$ is defined as 
\begin{equation}
\label{eq:chi2-stat}
\chi^2_{\rm stat}(\lbrace \Theta, \xi, \zeta \rbrace) = 
\sum_i 2 \left( N_i (\lbrace \Theta, \xi, \zeta \rbrace) - O_i + 
O_i \ln \frac{O_i }{N_i (\lbrace \Theta, \xi, \zeta \rbrace) } \right) \, ,
\end{equation}
where $O_i$ stands for the simulated ``observed'' event rates in the $i$-th energy bin, which we always take to be those predicted in the standard three-family framework. On the other hand, $N_i (\lbrace \Theta, \xi, \zeta \rbrace)$ stands for the predicted event rates in the same bin for a new physics model which depends on a set of parameters generically denoted as $\lbrace\Theta\rbrace$. It also depends on the nuisance parameters, as: 
\begin{equation}
\label{eq:Nch}
N_i (\lbrace \Theta, \xi, \zeta \rbrace) = 
\sum_s (1 + \xi^{\rm sig}_{i} + \zeta_s) \, s_i (\lbrace{\Theta}\rbrace) + 
\sum_b (1 + \xi^{\rm bg}_{i} + \zeta_b) \, b_i (\lbrace{\Theta}\rbrace) \, ,
\end{equation}
where we have included a set of possible signal ($s$) and background ($b$) contributions to the event rate sample in the bin. Also, from Eq.~\eqref{eq:Nch} it is clear that new physics effects are included on the backgrounds event rates as well (and not just on the signal). 

A pull term is then added to Eq.~\eqref{eq:chi2-stat} for each nuisance parameter, and the final $\chi^2$ is obtained after minimization over all nuisance parameters included in the fit. The final $\chi^2$ then reads:
\begin{align}
\chi^2_{\rm min} (\lbrace\Theta\rbrace) = {\rm min}_{\lbrace \xi, \zeta \rbrace } &
\left[ \chi^2_{\rm stat}(\lbrace \Theta, \xi, \zeta \rbrace)  
+ \sum_s \left( \frac{\zeta_s}{\sigma_{{\rm norm}, s}}\right)^2 
+ \sum_b \left( \frac{\zeta_b}{\sigma_{{\rm norm}, b}}\right)^2 \right. \nonumber   \\ 
& \left. + \sum_{i} \left( \frac{\xi^{\rm sig}_{ i}}{\sigma_{\rm shape, sig}}\right)^2 
+ \sum_{i} \left( \frac{\xi^{\rm bg}_{i}}{\sigma_{\rm shape, bg}}\right)^2 \right] \, ,
\label{eq:chi2min}
\end{align}
where $\sigma_{{\rm norm}, c}$ indicates the prior uncertainties assumed for the normalization of each signal or background contribution, while $\sigma_{\rm shape}$ stands for the shape uncertainties assumed for the overall background or signal event rates in the sample considered. 

A series of simplifications and approximations have been performed in order to keep the number of nuisance parameters as low as possible and to avoid numerical issues during minimization. First, we assume that the shape uncertainties are the same in each energy bin and therefore have removed the index $i$ in the corresponding priors as can be seen from Eq.~\eqref{eq:Nch} and~\eqref{eq:chi2min}. Next, we have assumed that all normalization uncertainties are uncorrelated between different signal and background contributions as well as between different event samples. Although in practice some degree of correlation is expected, our approach avoids numerical difficulties during minimization, which is important given the large number of nuisance parameters involved in the fit. It is also a conservative approach, as including correlations will most likely lead to an effective reduction on the size of systematic errors and a corresponding increase in sensitivity. Finally, although in the most general case one should in principle assume that both the signal and background would be affected by shape uncertainties, this would lead to a large number of nuisance parameters in the fit. Therefore, shape uncertainties are only included for the background contributions to the $\nu_e$- and $\nu_\tau$-like samples, while for the $\nu_\mu$-like sample we do include it both for signal and background rates. With the aim of keeping the simulations feasible, we include shape uncertainties for the sum of all background contributions in all cases, while we do include a separate normalization uncertainty for each contribution separately. In this way, we allow the different background contributions to float independently in the fit, while the shape systematics would affect the overall background rates. We believe this is enough given that in any case the background is typically dominated by a single channel (intrinsic contamination in the case of $\nu_e$ and $\bar\nu_e$ samples, or NC events in the case of $\nu_\mu$ and $\nu_\tau$-like samples, see Tab.~\ref{tab:events}). The values assumed for the prior uncertainties are summarized in Tab.~\ref{tab:sys}.

As shown in Sec.~\ref{sec:results}, in the case of sterile neutrino oscillations in the $\nu_\mu \to \nu_e$ channel, it is convenient to add information from the $\nu_e \to \nu_e$ and $\nu_\mu \to \nu_\mu$ event rates as well. While information from $\nu_e \to \nu_e$ oscillations is already included in the $e$-like event sample (since the intrinsic $\nu_e$ background event rates would be affected by oscillations), it is necessary to combine this information with the one extracted from the $\mu$-like event rates. In this case, a separate $\chi^2$ contribution is computed for the two samples using Eq.~\eqref{eq:chi2min}, and the total $\chi^2$ is obtained after adding up the two.

\section{Numerical implementation of oscillations in the average-out regime}
\label{app:average-out}

GLoBES uses a sampling method to evaluate the number of events in each bin. 
To avoid fast oscillations leading to aliasing, a low-pass Gaussian filter is available~\cite{Huber:2004ka,Huber:2007ji}. However, by default this filter does not handle well sampling points\footnote{In GLoBES, the term ``sampling points'' refers to the values of the true neutrino energies that are used in the computation of the event rates. See the GLoBES manual for additional details. } which are not equally separated. This is relevant for our DUNE implementation, due to the very different bin sizes used at low and high energies. In this appendix we present a formal description of the Gaussian filter, and we describe the way it has been implemented in our code to avoid this issue. 

Our goal is to perform the average of the probability within each sampling bin. We will consider that the value of the $E_i$ in each sampling bin follows a normal distribution with mean given by the central value of the bin ($E^c_i$) and standard deviation given by the size of the bin ($\Delta E_i$). The energy enters in the oscillatory part of the probability as $ \sin^2(\frac{\Delta m^2_{41}L}{4E})$, therefore to perform the average we need to know the probability distribution of  $\frac{L}{E}$. This leads us to the next formal problem.

Let us consider two independent random variables, normally distributed, $X \sim \mathcal{N}\left(\mu_{x}, \sigma_{x}^{2}\right)$ and $Y \sim \mathcal{N}\left(\mu_{y}, \sigma_{y}^{2}\right)$. Then the ratio $Z = \frac{X}{Y}$ follows the non-trivial distribution~\cite{twonormals}: 
\begin{equation}
\label{Ratio_distribution}
\small
f_{Z}\left(z ; \beta, \rho, \delta_{y}\right)=\frac{\rho}{\pi\left(1+\rho^{2} z^{2}\right)}\left\{\exp \left[-\frac{\rho^{2} \beta^{2}+1}{2 \delta_{y}^{2}}\right]+\sqrt{\frac{\pi}{2}} q \operatorname{erf}\left(\frac{q}{\sqrt{2}}\right) \exp \left[-\frac{\rho^{2}(z-\beta)^{2}}{2 \delta_{y}^{2}\left(1+\rho^{2} z^{2}\right)}\right]\right\} \, ,
\end{equation}
where 
\begin{eqnarray}
q & = &\frac{1+\beta \rho^{2} z}{\delta_{y} \sqrt{1+\rho^{2} z^{2}}} \, , \\
\beta &= &\frac{\mu_{x} }{ \mu_{y}} \, , \\
\rho & =&\frac{\sigma_{y} }{ \sigma_{x}} \, , \\
\delta_y &=& \frac{\sigma_{y} }{ \mu_{y} }\, . 
\end{eqnarray}
However, when $\delta_y < 0.1$ the density distribution in Eq.~\eqref{Ratio_distribution} also follows a normal distribution to a very good approximation, namely $\frac{X}{Y} \sim \mathcal{N} \left(\mu_x /\mu_y, \sigma_{X/Y}^{2}\right)$, where:
\begin{equation}
\sigma_{X / Y}^{2}=\left(\frac{\sigma_y }{\mu_y}\right)^{2} \left[\left(\frac{\sigma_x }{\sigma_y}\right)^{2} 
+ \left(\frac{\mu_x }{\mu_y} \right)^{2}\right] \, .
\end{equation}

In our case, $Y = L$ and $X = E_i$, and therefore $\mu_x = L$, $\mu_y = E^{\text{c}}_i$, $\sigma_x = 0$, $\sigma_y = \Delta E_i$. The condition that allow us to approximate the distribution of $L/E_i$ by a normal distribution is now given by $\Delta E_{i}<0.1 E^{\text{c}}_{i}$, implying that the choice of the sampling bins cannot be arbitrary if we want to use a Gaussian filter. Under these conditions the density distribution for $L/E_i$ is given by:
\begin{equation}
f_{L / E_{i}}\left(\frac{L}{E}\right) \approx \frac{1}{\sigma_{L / E_i} \sqrt{2 \pi}} e^{-\frac{1}{2}\left(\frac{L / E- L / E^{\text{c}}_{i}}{\sigma_{L / E_{i}}}\right)^{2}} \, ,\; \mathrm{with} \; \sigma_{L / E_i} = L\frac{\Delta E_{i} }{ (E^c_{i})^{2}} \, .
\end{equation}
Therefore in each sampling bin the average probability is given by:
\begin{equation}\label{Prob_average}
\left\langle P_{\alpha \beta}(L / E_i)\right\rangle = 
\int_{0}^{\infty} P_{\alpha\beta}(L/E_i) f_{L/E_i}\left(\frac{L}{E}\right) d\frac{L}{E} \, .
\end{equation}
In fact, the result from this expression can be computed analytically~\cite{Giunti:2007ry}, yielding the following result for the case of the sterile oscillation probability at the ND:
\begin{equation}
\label{Sterile_average}
\small
\left\langle P_{\alpha \beta}(L / E_i)\right\rangle=\delta_{\alpha \beta}-4\left|\mathcal{U}_{\alpha 4}\right|^{2}\left(\delta_{\alpha \beta}-\left|\mathcal{U}_{\beta 4}\right|^{2}\right)\left[\frac{1}{2}-\frac{1}{2} \cos \left(\frac{\Delta m_{41}^{2}L}{2E^{c}_{i}}\right) \exp \left\lbrace-\frac{1}{2}\left(\frac{\Delta m_{41}^{2} \sigma_{L / E_i}}{2}\right)^{2}\right\rbrace\right]
\end{equation}
which is the expression we use for our sterile neutrino analysis with GLoBES.

\bibliographystyle{JHEP}
\bibliography{references}

\end{document}